\documentclass[10pt,letterpaper]{article}
\usepackage[top=0.85in,left=2.75in,footskip=0.75in]{geometry}

\usepackage[caption = false]{subfig}

\usepackage{changepage}

\usepackage[utf8]{inputenc}

\usepackage{textcomp,marvosym}

\usepackage{fixltx2e}

\usepackage{amsmath,amssymb}

\usepackage{cite}

\usepackage{nameref,hyperref}

\usepackage[right]{lineno}

\usepackage{microtype}
\DisableLigatures[f]{encoding = *, family = * }

\usepackage{rotating}

\raggedright
\usepackage[aboveskip=1pt,labelfont=bf,labelsep=period,justification=raggedright,singlelinecheck=off]{caption}

\makeatletter
\renewcommand{\@biblabel}[1]{\quad#1.}
\makeatother

\date{}

\usepackage{lastpage,fancyhdr,graphicx}
\usepackage{epstopdf}
\fancyheadoffset[L]{2.25in}
\fancyfootoffset[L]{2.25in}
\usepackage{booktabs} % Horizontal rules in tables
\newcommand{\norm}[1]{\lVert#1\rVert}

\begin{document}
\vspace*{0.35in}

% Please capitalize all terms in the title except conjunctions, prepositions, and articles.
\begin{flushleft}
{\Large
\textbf\newline{Disease Spread through Animal Movements: A Static and Temporal Network Analysis of Pig Trade in Germany}
}
\newline
\\
Hartmut H. K. Lentz\textsuperscript{1,*},
Andreas Koher\textsuperscript{2},
Philipp Hövel\textsuperscript{2, 3},
Jörn Gethmann\textsuperscript{1},
Carola Sauter-Louis\textsuperscript{1},
Thomas Selhorst\textsuperscript{4},
Franz J. Conraths\textsuperscript{1},
%with the Lorem Ipsum Consortium\textsuperscript{\textpilcrow}
\\
\bigskip
\bf{1} Institute of Epidemiology, Friedrich-Loeffler-Institute, Südufer 10, 17493 Greifswald, Germany
\\
\bf{2} Institute of Theoretical Physics, Technische Universit{\"a}t Berlin, Hardenbergstraße 36, 10623 Berlin, Germany
\\
\bf{3} Bernstein Center for Computational Neuroscience Berlin, Humboldt-Universit{\"a}t zu Berlin, Philippstra{\ss}e 13,
10115 Berlin, Germany
\\
\bf{4} Unit Epidemiology, Statistics and Mathematical Modelling, Federal Research Institute for Risk Assessment, Alt-Marienfelde 17-21, 12277 Berlin, Germany
\bigskip

% Insert additional author notes using the symbols described below. Insert symbol callouts after author names as necessary.
% 
% Remove or comment out the author notes below if they aren't used.
%
% Primary Equal Contribution Note
%\Yinyang These authors contributed equally to this work.

% Additional Equal Contribution Note
% Also use this double-dagger symbol for special authorship notes, such as senior authorship.
%\ddag These authors also contributed equally to this work.

%\textcurrency a Insert current address of first author with an address update
% \textcurrency b Insert current address of second author with an address update
% \textcurrency c Insert current address of third author with an address update


% Group/Consortium Author Note
%\textpilcrow Membership list can be found in the Acknowledgments section.

% Use the asterisk to denote corresponding authorship and provide email address in note below.
* Corresponding Author: hartmut.lentz@fli.bund.de

\end{flushleft}
\section*{Abstract}

\paragraph{Background.}
Animal trade plays an important role for the spread of infectious diseases in livestock populations.
As a case study, we consider pig trade in Germany, where trade actors (agricultural premises) form a complex network.
The central question is how infectious diseases can potentially spread within the system of trade contacts.
We address this question by analyzing the underlying network of animal movements.

\paragraph{Methodology/Findings.}
The considered pig trade dataset spans several years and is analyzed with respect to its potential to spread infectious diseases.
Focusing on measurements of network-topological properties, we avoid the usage of external parameters, since these properties are independent of specific pathogens.
They are on the contrary of great importance for understanding any general spreading process on this particular network.
We analyze the system using different network models, which include varying amounts of information: (i) static network, (ii) network as a time series of uncorrelated snapshots, (iii) temporal network, where causality is explicitly taken into account. 

\paragraph{Findings.}
Our approach provides a general framework for a topological-temporal characterization of livestock trade networks.
We find that a static network view captures many relevant aspects of the trade system, and premises can be classified into two clearly defined risk classes.
Moreover, our results allow for an efficient allocation strategy for intervention measures using centrality measures.
Data on trade volume does barely alter the results and is therefore of secondary importance.
Although a static network description yields useful results, the temporal resolution of data plays an outstanding role for an in-depth understanding of spreading processes.
This applies in particular for an accurate calculation of the maximum outbreak size. 

% Please keep the Author Summary between 150 and 200 words
% Use first person. PLOS ONE authors please skip this step. 
% Author Summary not valid for PLOS ONE submissions.   
%\section*{Author Summary}

% \linenumbers

\section*{Introduction}
Infectious diseases in livestock can spread via various paths.
One of the main transmission routes is livestock trade \cite{Buttner2013418,fevre_animal_2006,green_modelling_2006,ribbens_transmission_2004,thrusfield_veterinary_2007}.
Other transmission routes include direct contact, aerial transmission (e.g. geographical closeness to an index premise) \cite{Mayer:2008ip,Ducheyne:2007gp, MartinezLopez:2011db} and vectors (insect, human, appliances) \cite{Olofsson:2014gk,Wang2012543}.

Livestock trade is of particular importance, since infectious animals can transmit a disease over long distances between premises.
Therefore, massive trade restrictions are implemented in case of an outbreak of a highly contagious disease such as classical swine fever \cite{oie:2015_code}.
However, before the first disease case is detected, the disease can spread unrestrictedly via trade.
The timespan of this unrestricted trade is called high risk period and can take weeks to months \cite{KnightJones2013161, Fritzemeier:2000p5909}.
In addition, trade restrictions are normally not implemented for some endemic diseases such as salmonellosis.
Hence, these diseases might freely spread via trade.

The spreading of an infectious disease by trade involves a number of different actors (e.g. farms, slaughterhouses or traders).
As these actors form a complex trade system, it is crucial to have an understanding of this trade system.
Such systems can be modeled as complex networks.

This analysis focusses on the German pig trade network as a substrate for spreading of infections in pigs.
The German pork industry is one of the largest in the world.
In the years 2011–2013, Germany was the third largest pork producer after the China and the USA \cite{faostat}.
About 4.5 million tons of pork meat are produced in Germany every year.
The gain in production value is about 7 billion Euros per year \cite{bmel:2015}.
For the classical swine fever outbreaks in Germany in the 1990s, it has been shown that the most frequent source of infection in secondary outbreaks was the trade with infected pigs \cite{Fritzemeier:2000p5909}.

It is the aim of this work to clarify how a disease principally can spread along the German pig trade network.
This means estimating the potential transmission ways of a disease between premises connected by direct or indirect trade contacts.
In other words, the considered network forms the basis (i.e. the substrate) for disease spread via pig trade in Germany.
In reality the spread of infectious disease via trade depends on additional parameters.
These parameters can be disease specific (e.g. virulence), farm specific (e.g. biosecurity level) or behavioral.
Since these specific parameters would bias the principal spreading pathways via trade, we exclude them from our analysis.
Hereby, the considered spreading mechanism mimics theoretically possible spreading paths – even if the true transmission probability might be lower for instance due to biosecurity measures.

In order to achieve the aim of revealing potential infection paths in the German pig trade network, the contact patterns between the actors of the system have to be analyzed.
In general, contact patterns among hosts forming a contact network are considered as one of the most critical factors contributing to inhomogeneous pathogen transmission, where the assumption of a mass-action process does not hold.
During the last decades, veterinary epidemiologists have been focusing on the disease transmission between livestock farms.
Premises and animal movements between premises can be translated into nodes and edges of a contact network, respectively.

Techniques adopted from social network analysis (SNA) have been intensively applied by veterinary epidemiologists in order to get a better understanding of the spatio-temporal livestock disease dynamics \cite{MartinezLopez2009, Dube:2009gx, Bigras:2007, bigras-poulin_network_2006, Buttner2013418, Christley:2005, Dutta:2014hx, Rautureau:2012, Webb:2005fs, Valdano:2015jt} and to identify network actors that are central to the spread of infectious diseases \cite{Natale:2009p6279, Natale:2011ch}.

In order to understand the dynamics of disease spread in complex networks, it is essential to analyze their large scale structure.
This is necessary to estimate the size and the incidence rate of a disease outbreak.
With this information the consequences of the introduction of a contagious disease can be estimated and control measures can be planned.
If nodes and edges differ from one another with respect to their centrality, i.e. to their potential to spread disease, this variability can be used to rank the nodes and edges.
Such a ranking allows veterinary authorities to select nodes and edges for the implementation of targeted surveillance and control measures following a central things first rule \cite{Stark:2006hw}.
Node rankings can be refined using meta information in form of edge weights.
In addition, using the temporal resolution of trade data provides a much more realistic picture of possible outbreak dynamics.

Concerning the German pig trade system as a complex network, there is no systematic characterization of this system in the literature so far.
Remarkable exceptions are \cite{Buttner2013418}, where a subset of the whole network was analyzed including different production types and \cite{Konschake:2013js}, where the German pig trade network was analyzed using a data-driven approach.
In this work we characterize the static and temporal network of pig trade in Germany as a substrate for spreading processes for the first time.

This article is an attempt to provide a comprehensive picture and characterization of the German pig trade network.
In order to give a transparent picture of the network, we hereby avoid the usage of external parameters whenever possible.
Therefore, neither explicit disease specific parameters nor specific intervention measures such as trade restrictions are considered.
In addition, the analysis provided here can be regarded as a general framework to investigate livestock trade networks.

This paper is structured as follows:
first, we briefly describe the data under consideration.
In Section \emph{Static Network Analysis} we analyze the pig trade data as a static network, where we characterize the network from a large scale perspective and discuss different strategies for targeted vaccination.
Section \emph{Weighted Network Analysis} gives a brief overview of the impact of trade volume on the static network results.
We consider the temporal resolution of the network data set in Sections \emph{Network as Time Series} and \emph{Temporal Network Analysis}.
The network is considered as a time series of uncorrelated snapshots in Section \emph{Network as Time Series}.
Finally, we take into account causality for network traversal in Section \emph{Temporal Network Analysis}.

\subsection*{Data}
\begin{figure}[htbp]
\begin{center}
\includegraphics[]{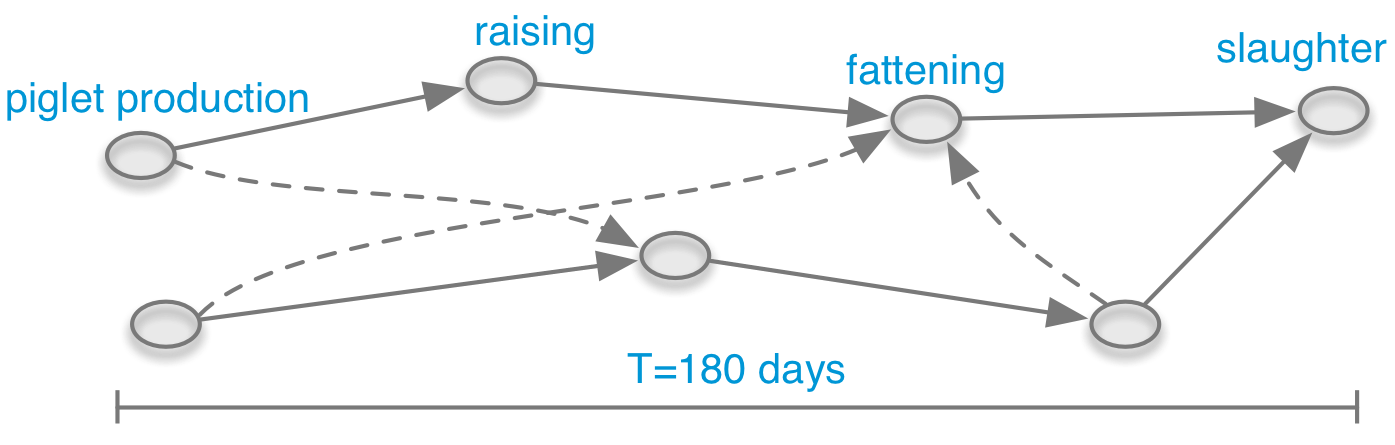}
\caption{Schematic of pork production chains (solid arrows) in the pig-trade network. Different production chains can be connected by additional cross links (dashed arrows).}
\label{fig:pork_prod_chain}
\end{center}
\end{figure}

In this article we analyze an extract of the HI-Tier database \cite{HI-Tier}.
The database comprises livestock movements of pigs in Germany since 2006.
The extract under consideration represents the trade between premises of the pork production chain in Germany in the period between 2011-01-01 and 2014-12-31.
Considered data are owner, prepossessor, trade date and trade volume.
Thereof a network is generated where trading premises are \emph{nodes} that are connected by directed \emph{edges} (trade links).
In addition, trade volume can be included as further information giving a weight to each edge of the network.
The system consists of elementary pork production chains.
Fig.~\ref{fig:pork_prod_chain} depicts a schematic illustration of the underlying farming system of the network.
The figure shows only the production chain of piglet production, raising, fattening and slaughter.
Traders and breeding are not shown.

\section*{Analysis}
The resulting network is analyzed in different representations:
\begin{enumerate}
\item \emph{Static network}.\label{en:uno}
Direction of trade is taken into account\footnote{In principle, the network could also be considered undirected.
However, this approach is less meaningful in this context due to the directed nature of the involved production chain.}.
Trade connections are aggregated over time, i.e. the network is static.
A trade link is drawn if there is at least one trade action over the observation period.
In addition, we analyze the impact of trade volume.
\item \emph{Network as time series}. \label{en:dos}
The system is considered as a time series of directed network snapshots at different time steps.
Edge weights are considered to some extent.
\item \emph{Temporal network}. \label{en:tres}
The system is considered as a time series of directed network snapshots at different time steps.
In addition, causality is fully considered for network traversal via edge sequences.
\end{enumerate}
In each case the network comprises of 97,980 nodes.
The static representation (\ref{en:uno}.) consists of 315,333 edges.
For the temporal cases (\ref{en:dos}.) and (\ref{en:tres}.) the data set contains 6,359,697 trade transactions (edges).
The observation period is $T = 1461~ d$, i.e. 1.5 million edges per year.

\subsection*{Static Network Analysis}%\label{sec:static}
In this section, we analyze the trade data as a static network.
A static network or \emph{graph} $G=(V,E)$ consists of a set of nodes $V$ and a set of edges $E$, where every edge connects a pair of nodes.
In the considered network, edges have a direction given by trade.
Mathematically, a network can be represented as an adjacency matrix $\mathbf{A}$ with elements $(\mathbf{A})_{ij} = 1$ if there is an edge from node $i$ to node $j$, and $(\mathbf{A})_{ij} =0$ otherwise.
The total number of neighbors (trade partners) of a node (premise) is called its \emph{degree} or \emph{total degree}.
If edge direction is considered, we distinguish between \emph{in-degree} (incoming edges) and \emph{out-degree} (outgoing edges).

\paragraph{Large Scale Structure I -- Components.}
In the following, we investigate the component structure of the static network by means of components.
We will see that the directed nature of trade plays a major role here.

In principle, the outbreak size of any epidemic on a network is limited by the component structure of the network as a worst case scenario.
A \emph{component} is a subset of nodes $C \subseteq V$ for which a path exists between any pair of nodes in $C$.
A \emph{path} $P_{i \rightarrow j}$ between two nodes $i$ and $j$ is an indirect connection between them via arbitrarily many edges without traversing a node twice.
Note that for directed networks, $P_{i \rightarrow j}$ does not necessarily imply $P_{j \rightarrow i}$.
In general there may exist a large number of paths between two nodes.
In this article, by path we always mean the \emph{shortest path}, i.e. the $P_{i \rightarrow j}$ with the smallest number of traversed edges.
The average shortest path length in the considered network is 5.5, i.e. on average it takes 5.5 steps to go from a randomly chosen node to another randomly chosen node.
The maximum shortest path length is called \emph{diameter} and its value is 18 for the considered network (see Table ~\ref{tab:properties}).

Neglecting the directionality of edges, the network exhibits a giant component, which in directed network is commonly called giant weakly connected component (GWCC).
For the considered network, it comprises almost all nodes (see Table \ref{tab:properties}).
This means that virtually all nodes of the network are at least 'touched' by trade connections.
We find that 99 \% of all nodes are connected through trade contacts.
Nodes not belonging to the GWCC form other components which are only very small islands in the network.
\begin{table}[htp]
\caption{Standard network properties of the static German pig trade network. Diameter and shortest path length are computed for the GSCC.}
\begin{center}
\begin{tabular}{lr}
\toprule
\textbf{Property} & \textbf{Value}\\
\midrule
Number of nodes & 97,980\\
Number of edges & 315,333\\
Edge density & $3.2 \times 10^{-5}$\\
\hline
size of GSCC & 28 \%\\
size of GWCC & 99 \%\\
%link reciprocity $\varphi _\text{link}$ & $0.114 \pm 0.002$\\
\hline
diameter & 18\\
av. shortest path length & 5.5\\
\hline
path density & 0.24 \\
\bottomrule
\end{tabular}
\end{center}
\label{tab:properties}
\end{table}%

The formation of a giant component is also known as percolation and has been studied in a variety of systems \cite{RevModPhys.74,Dorogovtsev:2001jd,Newman2003}.
From the statistical point of view, a giant component emerges if the number of edges exceeds a certain threshold \cite{Newman2003}.

\begin{figure}[htbp]
\begin{center}
\includegraphics[]{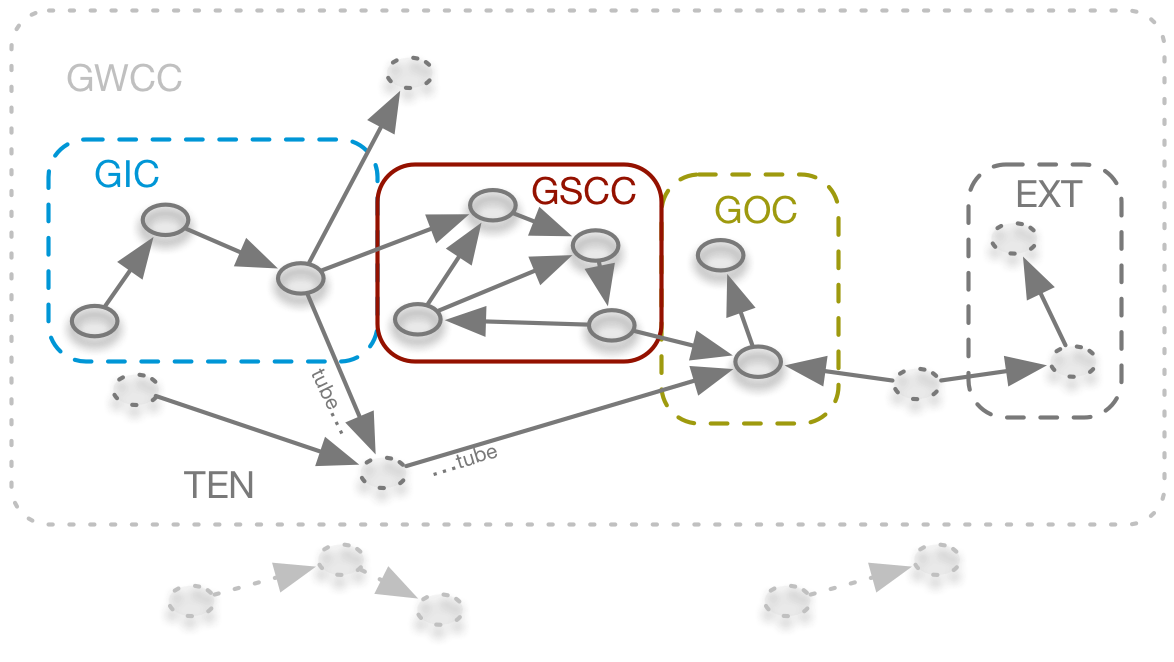}
\caption{Component structure of directed networks. The giant strongly connected component (GSCC) forms the center of the network (red box). Nodes of the GSCC and the giant in-component (GIC, dashed blue box) have a high spreading potential, whereas all other nodes (GOC - giant out component, dashed yellow box; TEN - tendril; EXT - external nodes, grey dashed box) cannot reach a significant fraction of the network. Box sizes do not reflect the actual sizes of the components. The giant weekly connected component (GWCC) is given by the grey dotted box.}
\label{fig:component_structure}
\end{center}
\end{figure}

If edge \emph{direction} is taken into account, the network shows a more complex component structure.
This is due to the reciprocity that is not guaranteed in directed networks.
The general structure of directed networks has been investigated in \cite{Dorogovtsev:2001jd}.
It is schematically depicted in Fig.~\ref{fig:component_structure}.
In analogy to the GWCC, the \emph{giant strongly connected component} (GSCC) is a subset of nodes for which a \emph{directed} path exists between all pairs of them.
For the considered data set, the GSCC contains about 1/4 of all nodes (red box in Fig.~\ref{fig:component_structure}) and forms the backbone of the network in the sense that it ensures the global connectivity of the network.
All nodes with access to the GSCC that are not themselves part of it, form the \emph{giant in-component} (GIC, blue frame in Fig.~\ref{fig:component_structure}).
In analogy, the \emph{giant out-component} (GOC) contains the nodes that can be reached from the GSCC, but are not part of the GSCC themselves (yellow box in Fig.~\ref{fig:component_structure}).

Besides the mentioned components, a directed network generally contains so-called \emph{tendrils} (TEN, grey dashed nodes in Fig.~\ref{fig:component_structure}).
Tendrils are sets of nodes that do not belong to the GSCC, but are reachable from the GIC or that can reach the GOC.
A special case of tendrils are \emph{tubes}, which start at the GIC and bypass the GSCC to end in the GOC.
Finally, \emph{external node sets} (EXT in Fig.~\ref{fig:component_structure}) are part of the GWCC, but have no access to the GOC.

\begin{figure}[htbp]
   \centering
   \includegraphics[]{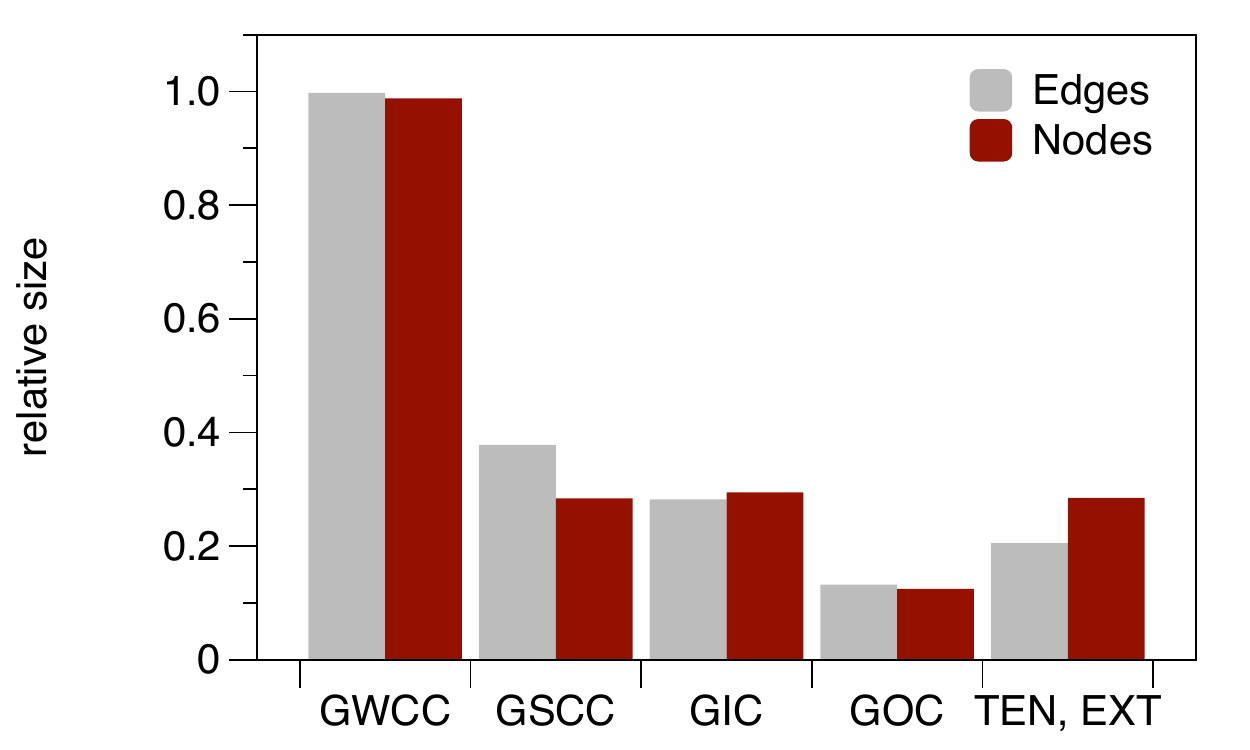} % requires the graphicx package
   \caption{Relative sizes of the large scale components of the pig trade network. Sizes are normalized to the total number of nodes in the network.}
   \label{fig:component_distr}
\end{figure}

Figure \ref{fig:component_distr} shows the distribution of nodes and edges of the different giant structures.
About half of the nodes are in the GSCC and the GIC.
They form the part of the nodes that can in principle cause larger outbreaks.
It is remarkable that the GOC makes up only a small part of the network.

The topology shown in Fig.~\ref{fig:component_structure} results in a salient property regarding the spreading potential of the nodes in the network.
The spreading potential of a node $i$ can be quantified using the number of nodes reachable from node $i$ by a path of arbitrary length.
We call this number the \emph{range} of node $i$ \cite{Lentz:2012pre}.
Therefore, the range of a node defines a simple measure for assessing the risk of a pathogen to spread via the network.

Figure \ref{fig:ranges_static} depicts the range of every node in the pig trade network.
The distribution shows a strong bimodal structure with two node classes:
(i) a class with long-range nodes and (ii) a class with short-range nodes.
This distribution can be explained with the component structure as described above:
\begin{figure}[htbp]
   \includegraphics[]{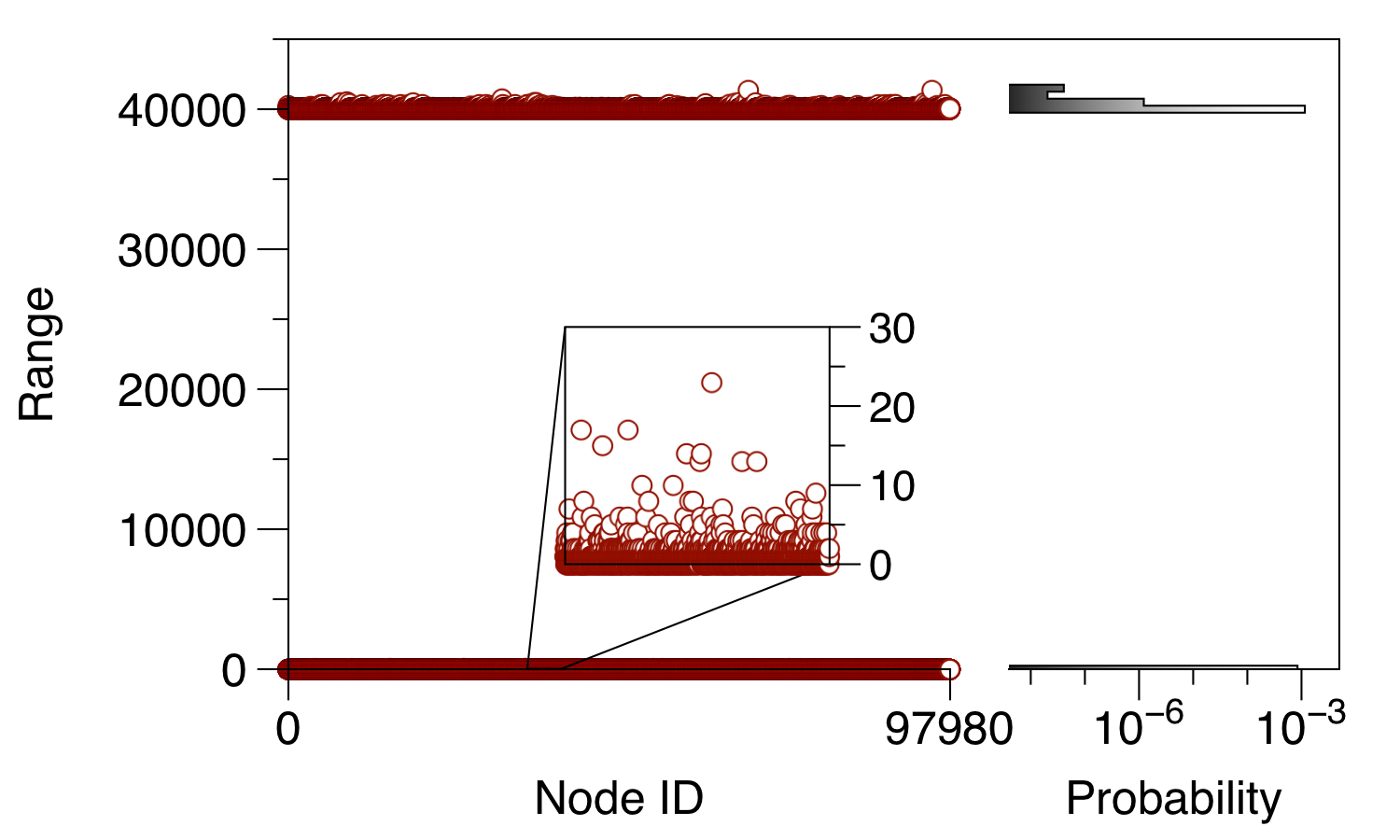} % requires the graphicx package
   \caption{Range for every node in the pig trade network. About 50 \% of the nodes have a long range of approximately 40,000 nodes, i.e. an infection started here could reach almost half the network. The other 50 \% of the network nodes have a short range and cannot cause large outbreaks (maximum short range: 70). The probability distribution of the ranges is shown in the right panel.}
   \label{fig:ranges_static}
\end{figure}
The nodes belonging to the GSCC or GIC have a long range.
They make up a fraction of 58 \% (56,656 nodes).

All other nodes (GOC, TEN, EXT) show a considerably shorter range.
They represent a much smaller risk for the spread of infectious diseases.
On the other hand, the maximum range of all short range nodes is still 70.% (inset in Fig.~\ref{fig:ranges_static}).

In principle, tendrils might form large structures as well.
The size of these structures can be estimated by removing the GSCC from the network and computing the ranges of the remaining nodes.
The maximum range of the tendril nodes is 48 for the pig-trade network.
Nodes of the GOC have a maximum range of 40, whereas EXT nodes can reach up to 70 nodes. 
Overall, this shows that disease spread via trade in GOC, TEN and EXT cannot cause large outbreaks.
Table~\ref{tab:maxranges} shows the maximum ranges in the different giant structures.
\begin{table}[htp]
\caption{Maximum ranges for the large scale structures of the network.}
\begin{center}
\begin{tabular}{llr}
\toprule
\textbf{Component}& \textbf{Category} & \textbf{Max. range}\\
\midrule
GIC & long range & 41,369\\
GSCC& long range & 40,040\\
\hline
GOC& short range & 40\\
TEN& short range & 48\\
EXT& short range & 70\\
\bottomrule
\end{tabular}
\end{center}
\label{tab:maxranges}
\end{table}%

It should be noted that the observed range behavior (Fig.~\ref{fig:ranges_static}) is typical for directed networks.
Since (livestock) trade networks are generally directed, a similar behavior can be expected also for other trade networks.
Contrary to trade networks, \emph{social} networks, which can be used to model local spreading dynamics, are rather undirected and do not reveal the features shown in Figures~\ref{fig:component_structure} and \ref{fig:ranges_static}.

\paragraph{Large Scale Structure II -- Mixing Patterns.}
In the absence of information on the internal contact structure within a population, a widely used assumption is homogenous mixing.
This means that every node could be in contact with any other node in the network and there is no selection bias or characteristic interaction pattern.
In this section, we will quantify the strength of mixing structures as they are contained in the pig trade data set.
These structures can be arbitrary node categories that have to be defined in the first place (e.g. node type, administrative regions, premise size, hygiene status, etc.).
In the following we restrict our analysis to categories that are intrinsically linked to the data at hand.
These are:

\begin{enumerate}
\item Federal state
\item District
\item Municipality
\item Degree (number of trade links).
\end{enumerate}

First, we assign a category $k$ to every node in the network, e.g. the federal state in which the node (premise) is located.
Then, we compute the number of edges between nodes of the categories (e.g. federal states) $k$ and $l$.
This can be summarized in a mixing matrix $\mathbf{e}$ with elements \cite{Newman:2003p4336}
\begin{align}
(\mathbf{e})_{kl} = ~ &\text{fraction of links between} \nonumber \\
&\text{categories }k\text{ and }l \label{eq:def_e},
\end{align}
where the indices $k$ and $l$ represent different node categories.

The propensity of a network to prefer links between nodes of the same category can be quantified using the assortativity coefficient $\varphi $.
We first focus on the assortativity coefficient for enumerative node categories such as Federal state, District or Municipality.
The degree being a scalar node property is discussed below.
The \emph{enumerative assortativity coefficient} is defined as follows \cite{Newman:2003p4336}:
\begin{equation}\label{eq:cat_assort_coeff}
\varphi = \frac{\mathrm{Tr}~\mathbf{e} - \norm{\mathbf{e}^2}}{1-\norm{\mathbf{e}^2}},
\end{equation}
where $\mathrm{Tr}~\mathbf{e} =\sum _{i=j} e_{ij} $ is the trace of matrix $\mathbf{e}$ and $\norm{\mathbf{\cdot} }$ is the sum over all matrix elements.
Formally, $\varphi $ is a correlation coefficient.
If a network exhibits a positive assortativity coefficient $\varphi > 0$, it is called an assortative network (with respect to that category);
for $\varphi < 0$ the network is called disassortative.
Networks where $\varphi = 0$ are called uncorrelated.

In a perfectly assortative network ($\varphi =1$) all nodes are connected only to nodes of the same category, e.g. links are only formed between nodes of the same federal state or the same degree.
This would correspond to a mixing matrix $\mathbf{e}$ with finite elements only along the main diagonal, all other elements being zero.
On the other hand, the (enumerative) assortativity coefficient for a perfectly disassortative network is in general greater than $-1$ and the exact value depends on the number of considered categories \cite{Newman:2003p4336}.
%a perfectly disassortative network is in general closer to a random network, since applying a random rewiring of nodes for a given network, nodes of different categories are likely to be connected \cite{Newman:2003p4336}.
%This is particularly true, if many categories are present.
%Therefore, even strongly disassortative networks typically show values values $-1 < \varphi < 0$ closer to zero even in the case of a perfectly disassortative network.

Table~\ref{tab:mixing} shows the assortativity coefficients for the four node categories mentioned above.
The membership to a federal state represents a large scale classification for each node and the corresponding assortativity coefficient ($\varphi = 0.81$) is relatively high.
Premises have thus a preference to trade within the same federal state.
As a consequence, imposing trade restrictions along the borders of federal states results in a rather small modification of the original network in case of an outbreak \cite{Lentz:2011}.
Even though trade restrictions along the borders of federal states would disconnect the network, most trade connections are not affected, since the majority of trade connections connects node pairs in the same federal states.

\begin{figure}[htbp]
\begin{center}
\includegraphics[]{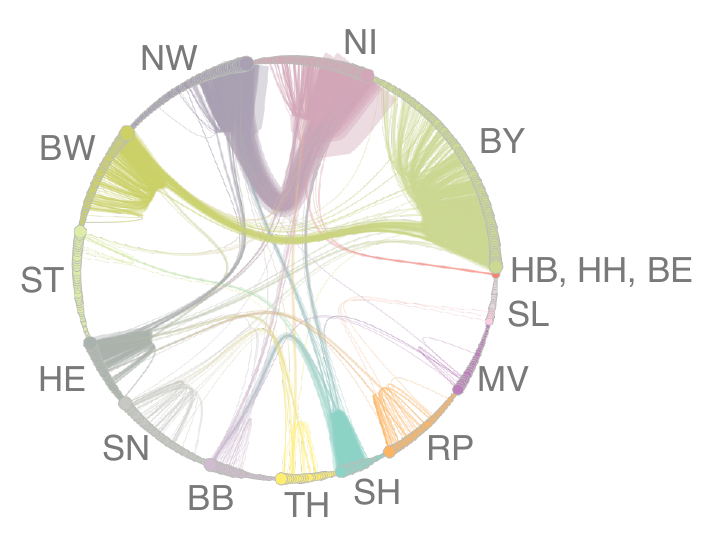}
\caption{Trade between districts in Germany. Node sizes correspond to the degree. Edges are bundled with respect to the federal states. Trade is dominated by links between districts in NW and NI and BY and BW, respectively.
Self-loops (intra-state trade) are not shown.}
\label{fig:edge_weighted_bundeslaender}
\end{center}
\end{figure}

Fig.~\ref{fig:edge_weighted_bundeslaender} shows the contact structure between the federal states in Germany.
Nodes are districts that are color-coded according to their federal state membership.
Intra-district trade links (self loops) are not shown.
Edge widths correspond to the number of trade links between two districts.
Node sizes correspond to the node degree.
For an improved visualization, edges are bundled for each federal state using the algorithm proposed in \cite{holten:2006}.

We observe again that the majority of trade takes place within the federal states (intra-district trade links not shown in Fig.~\ref{fig:edge_weighted_bundeslaender}).
Inter-state links are mainly formed between North Rhine-Westphalia (NW) and Lower Saxony (NI) as well as Bavaria (BY) and Baden-Wuerttemberg (BW).
In fact, the trade between NW and NI alone accounts for 36 \% of all inter-state trade connections. 
The federal states NI, NW, BW und BY account for 47 \% of all inter-state edges.
%The distribution of inter-state trade edges follows the Pareto-Principle, i.e. 20.5 \% of the districts contain 80.4 \% of the trade links.
Considering the districts of all federal states, 20.5 \% of the districts contain 80.4 \% of the inter-state trade links.

Contrary to federal states, there is less tendency that premises trade within their district or municipality (see Table~\ref{tab:mixing_SI}).
This may originate from the fact that not all types of premises needed for pork production are present in all districts or municipalities.
\begin{table}[htp]
\caption{Assortativity coefficients $\varphi $ of the static German pig trade network. The network is assortative with respect to federal state, district and municipality and disassortative with respect to node degree. $\sigma $ is a statistical error estimate. Assortativity coefficients for node categories are computed using Eq.~\eqref{eq:cat_assort_coeff}, for the degree Eq.~\eqref{eq:deg_assort_coeff} has been used.}
\begin{center}
\begin{tabular}{lrr}
\toprule
\textbf{Category} & $\varphi$ & $\sigma $\\
\midrule
Federal state & 0.81 & $7\times 10^{-4}$ \\
District & 0.43 &$2\times 10^{-4}$ \\
Municipality & 0.14 & $6\times 10^{-5}$\\
\hline
Degree & -0.13 & $1\times 10^{-3}$ \\
\bottomrule
\end{tabular}
\end{center}
\label{tab:mixing_SI}
\end{table}%

We now focus on the assortativity with respect to the node degree.
It is insightful to investigate whether or not nodes show a tendency to connect with nodes of similar degree.
If for example nodes of high in-degree preferably trade with nodes of high out-degree (disassortative network), disease spread would be (at least locally) facilitated.
It has been shown that targeted vaccination has a stronger effect in disassortative networks than in uncorrelated or assortative ones \cite{Newman:2002p4545}.

In contrast to the enumerative categories discussed above, the degree is a scalar assigned to every node in the network.
This implies that not only node pairs with exactly the same degree contribute to the degree assortativity coefficient, but also node pairs of similar degree.
Degree assortativity is therefore not analyzed using Eq.~\eqref{eq:cat_assort_coeff}, but is rather measured in a different way.
Following \cite{Newman:2003p4336}, the \emph{degree assortativity coefficient} is given by a Pearson correlation coefficient
\begin{equation} \label{eq:deg_assort_coeff}
\varphi = \frac{\sum _{xy} xy (e_{xy} -a_x b_y)}{\sigma _a \sigma _b} 
\end{equation}
where $e_{xy}$ are the entries of a mixing matrix containing the fraction of edges connecting nodes of degree $x$ and $y$, $a_x = \sum _y e_{xy}$ and $b_y = \sum _x e_{xy}$.
$\sigma _a$ and $\sigma _b$ are the standard deviations of $a$ and $b$, respectively.

The pig trade network is disassortative with respect to the degree.
Table~\ref{tab:mixing_SI} shows the assortativity coefficient for the total degree, which is smaller than zero.
Therefore, there is a tendency that nodes of different degree are connected.
This behavior is typical for technological and biological networks \cite{Newman:2002p4545}.
In our context, the disassortative degree mixing can be explained firstly by the fact that slaughterhouses receive animals from a large number of different farms (large degree), including many small ones (small degree) (see Fig.~\ref{fig:pork_prod_chain}).
A similar pattern is formed for piglet production, where few piglet producers provide piglets for a large number of different farms.
It is well known that the number of piglet-producers and slaughterhouses is small compared to the rest of the system \cite{eu_stat_food, stat_jahrbuch}.

Applying vaccination strategies in slaughterhouses might be considered effective, but it can not protect the rest of the network.
On the other hand, vaccination of piglet producers is a clearly effective, but rather obvious and trivial strategy.
In order to get an estimate for the center of the production chain, we compute the degree assortativity coefficient for the subnetwork where all nodes with vanishing in degree or out degree are removed.
Although this procedure is not exact, it should remove most slaughterhouses and piglet producers from the network.
The assortativity coefficient for this subnetwork is $-0.16$ and thus this part of the network is still disassortative with respect to the degree.

Besides the total degree, we also compute $\varphi $ for all combinations of in-degree and out-degree.
The values are similar to the total degree case and are shown in \nameref{S3}.

The standard deviation $\sigma $ of the mixing coefficient can be estimated statistically \cite{Newman:2003p4336}.
For the considered network all statistical errors are orders of magnitude smaller than the assortativity coefficients (see Table \ref{tab:mixing_SI}).
Thus, the observed results are highly statistically significant.

In conclusion, it follows from the analysis of mixing patterns that federal states provide an intrinsic partition of the network, even if this partition is large-scale.
In this context, also several federal states could be combined into even larger clusters.
Concerning the local mixing structure of the network, we expect that an efficient targeted vaccination is possible, since high-degree nodes tend to be connected to many low-degree nodes.
Hence, this analysis can contribute to define regions according to the OIE terrestrial animal health code \cite{oie:2015_code}.

\paragraph{Small Scale Structure - Centrality and Intervention Allocation.}
In order to implement efficient measures of disease control and surveillance, the infection risk of every node has to be assessed.
For this purpose, so-called centrality measures have been defined.
The range defined above is one such centrality measure.
We have already shown in Table~\ref{tab:maxranges} that the range allows us to identify two risk classes for the nodes in the network.

\begin{figure}[htbp]
   \centering
   \includegraphics[]{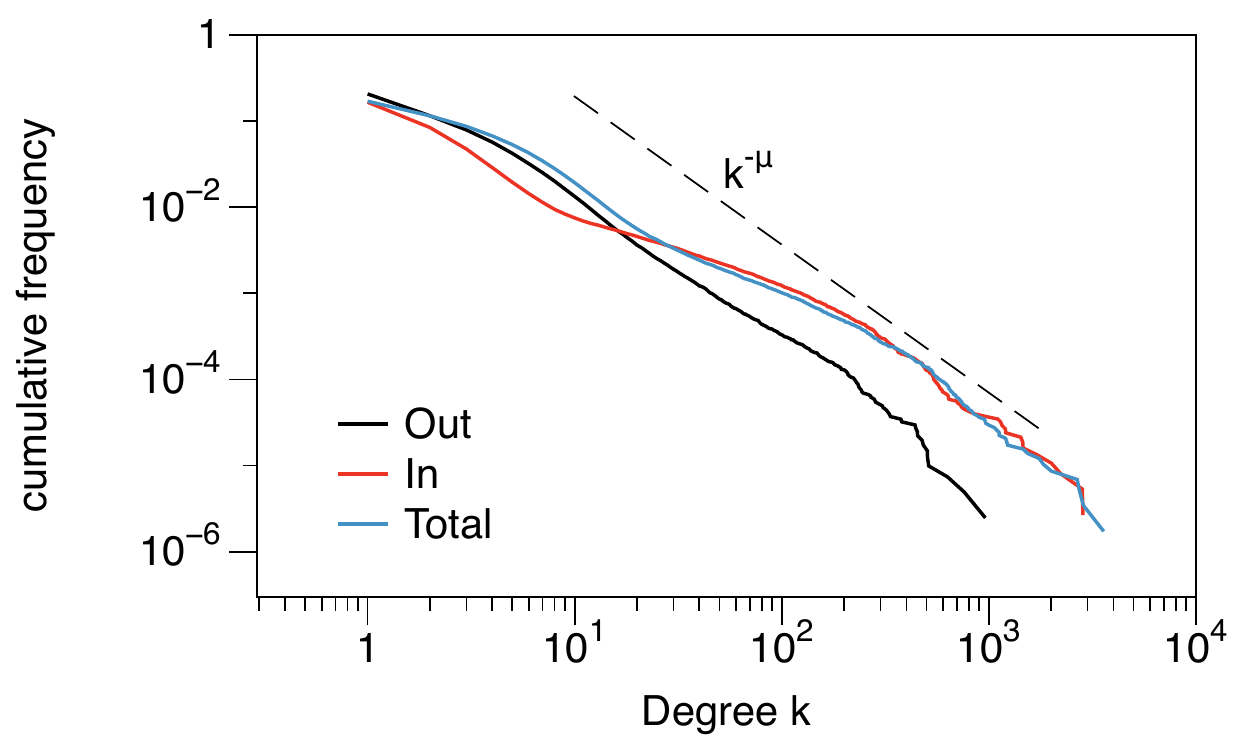} % requires the graphicx package
   \caption{Degree distribution of the livestock trade network. The out-degree distribution can be approximated by a power law of the form $p_k \sim k^{-\mu}$ with $\mu \approx 2.7$ (estimated using a maximum likelihood approach \cite{Clauset:2009}). The figure shows the cumulative distribution to minimize fluctuations.}
   \label{fig:degree_cdf}
\end{figure}

The simplest centrality measure is the \emph{degree} of a node, which can be easily obtained by counting the number of neighbors. 
In the case of the pig trade network, we distinguish between in-degree and out-degree; the total number of links a node is connected to is the total degree $k$.
Fig.~\ref{fig:degree_cdf} shows the degree distributions of the network.
The figure shows the cumulative distribution on log-scale.
The distributions are heavy tailed and the out-degree distribution can be approximated by a power-law, i.e. the distribution has the asymptotic form $p_k \sim k^{-\mu}$ with some constant $\mu$.
On the other hand the in-degree distribution exhibits a bimodal structure reflecting the existence of large slaughter houses.
It has been shown that the degree distribution has a significant impact on disease dynamics \cite{pastor-sat_2, Pastor-Satorras_vespi:2001, Albert:2000, Danon:2011hl, Lloyd:2001ud, Jones:2003p754, Kao:2006iz, Schwartz:2002dn, Meyers:2005p5782}.

In order to investigate different vaccination strategies for the network, the following centrality measures have been computed:
\begin{description}
\item[degree centrality.] $C_D$ -- Number of neighbors of a node. Normalized to the number of nodes in the network.
\item[betweenness centrality.] $C_B$ -- Frequency that a node lies on a shortest path between other nodes.
\item[closeness centrality.] $C_C$ -- Reciprocal average shortest path length between a node and all other nodes.
\end{description}
We also investigate other measures such as eigenvector centrality, pagerank und Katz centrality.
These measures, however, turn out to be less suited for disease control (see \nameref{S2}).
An overview of the role of different centrality measures for disease control is provided in \cite{Dube:2009gx} and \cite{MartinezLopez2009}.

Knowledge of the distribution centrality over the network can be used to implement targeted intervention measures.
For this purpose nodes are first of all ranked according to their centrality.
Then the impact of the removal of nodes with the highest rank on the functionality of the network can be measured.
After each node removal, centrality has to be computed again.
In our context, removing nodes does not necessarily mean that they are not active in the trade network, but rather that they effectively can not transmit a disease to other nodes.
This can be achieved by culling, isolation of animals, increased hygiene measures or vaccination.
The functionality of a network can be defined by the size of its GSCC, since the key feature of every network -- namely to ensure the interconnectedness between the nodes -- is manifested here.
If the size of the GSCC is reduced, the network disintegrates into smaller components and every disease outbreak is restricted to small 'islands'.

The impact of different vaccination strategies on networks has been analyzed in \cite{Albert:2000, holme:2002}.
The degree of a node has been shown to be a good indicator for its importance.
Furthermore, the degree is relatively easy to measure even if network data knowledge is limited.
In addition to the degree we study the suitability of the other centrality measures mentioned above for a risk ranking.

\begin{figure}[htbp]
   \centering
   \includegraphics[]{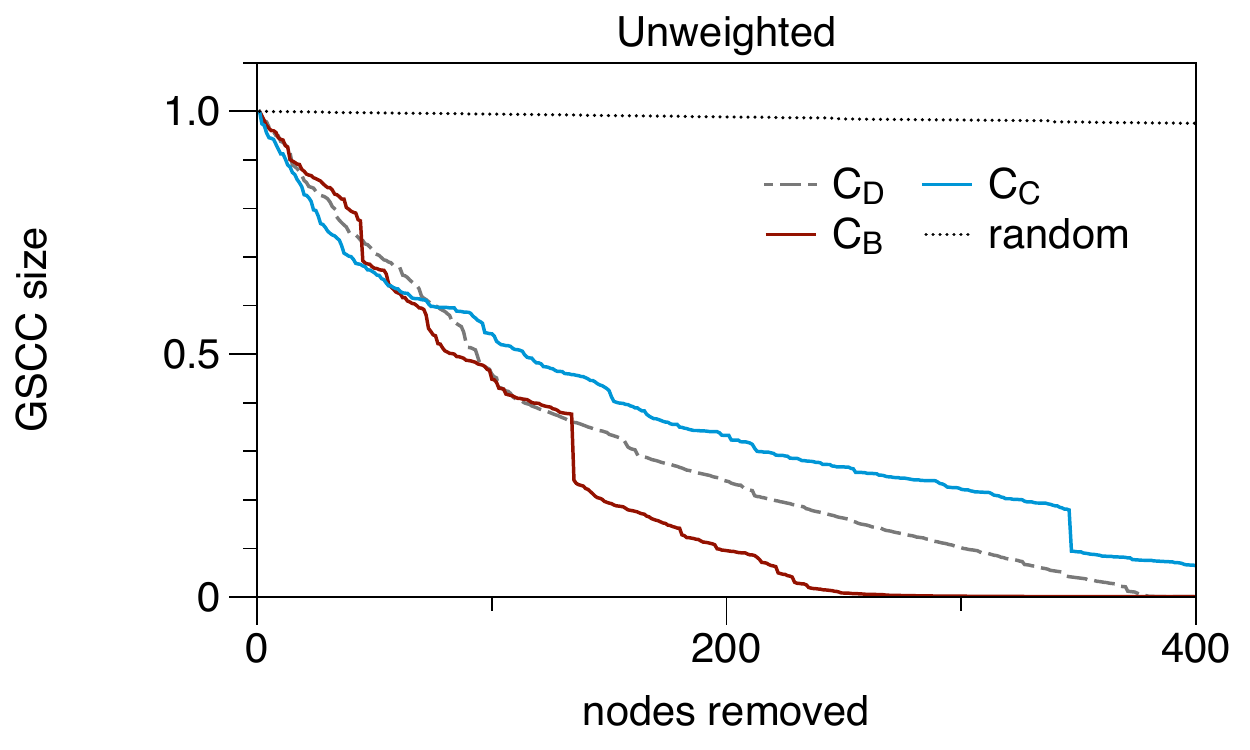} % requires the graphicx package
   \caption{Impact of centrality based node removal, when up to 1 \% of the nodes are removed. $C_D$-Degree centrality, $C_B$-Betweenness, $C_C$-Closeness. Size of giant strongly connected component is normalized to unity.}
   \label{fig:node_removal}
\end{figure}

Fig.~\ref{fig:node_removal} shows the size of the GSCC depending on the number of removed nodes;
nodes are removed according to their centrality rank in decreasing order.

It is remarkable that the removal of randomly chosen nodes barely has a measurable impact on the functionality of the network (dotted line in Figure \ref{fig:node_removal}).
This phenomenon has also been observed for other systems \cite{Albert:2000} and is related to the degree distribution of the network.
In the case of random removal, about 30,000 (of all 97,980) nodes must be removed in order to halve the size of the GSCC.
For comparison: in case of targeted removal of central nodes it suffices to remove only about 100 nodes to achieve the same effect.

It is apparent from Fig.~\ref{fig:node_removal} that the optimal strategy depends on the number of nodes to be removed.
In practice this number depends on the specifications of disease control.
Depending on the number of removed nodes, we define the optimal strategy as the one with the smallest value of the GSCC size at this point (Fig.~\ref{fig:node_removal}).
If about 50 nodes are to be removed from the network, nodes of high closeness or degree should be chosen.
In case of removing 100 or more nodes, nodes of high closeness are less efficient.
In this case, nodes of high degree and above all nodes with high betweenness should be removed.
Overall, betweenness centrality shows the best performance for disease control.

The large-scale structure of the network is also apparent in the distribution of centrality measures.
Nodes in the GIC tend to show a high out-degree and low in-degree, whereas the opposite holds for the GOC.
One could expect that many high in-degree nodes (slaughter houses) are located in the GOC.
Interestingly, nodes with the largest in-degrees are located in the GSCC.
These results are provided in \nameref{S3}.

We conclude:
\begin{enumerate}
\item any centrality based intervention performs significantly better than random intervention.
\item removal of high closeness or degree nodes is efficient for removal of up to 50 nodes.
\item removal of high betweenness nodes is efficient for removal of more than 100 nodes.
\end{enumerate}

\subsection*{Weighted Network Analysis}%\label{sec:weighted}
In principle, the number of traded animals plays an important role for the spread of animal diseases, since in reality hardly ever all animals in the outgoing premise are infectious.
Depending on the trade volume, this could have a strong impact on the epidemic conductivity.
Here we distinguish between the \emph{infection probability} and the \emph{infectiousness} of each edge in the network.

For highly contagious diseases (such as classical swine fever, Aujeszky’s disease, foot and mouth disease) a trade contact is even infectious, if a single infected animal transported.
Thus, the relevant measure here is the \emph{probability} that for a trade contact with another node at least one animal is infected.
This probability depends on the trade volume and the prevalence in the originating premise (see \nameref{S4}).
Given the relatively high trade volume in the network analyzed here, the mean transmission probability of a trade connection is close to 1 (see \nameref{S4}).

For lowly contagious diseases (such as Tuberculosis) the \emph{infectiousness} of a trade contact plays a central role.
Contrary to the infection probability discussed above the infectiousness of a link is closely related to the number of transported animals.
Thus, the link weight plays a central role here.

Although the weighted network is topologically similar to the unweighted network, there is evidence that the shortest path structure of the network is different for the weighted case.
In fact, we find an average shortest path length of $9.7$ and a diameter of $30$ for the weighted network.
Compared to the static network, both the average shortest path length and the diameter are twice as high (see Table~\ref{tab:properties}).
This implies that weighted shortest paths differ from purely topologically shortest paths.

Nevertheless, this circumstance does not seem to alter our findings for the unweighted network.
We analyze the weighted network with respect to targeted vaccination in \nameref{S4}.
The results found are qualitatively similar to the results of the previous section.
In brief, in the context of intervention measures, removing nodes with high weighted degree (i.e. trade volume) turns out to be an appropriate strategy.
Additionally, weighted closeness and weighted betweenness perform well as in the unweighted case.
Their performance is, however, not superior to the unweighted case. 

\subsection*{Network as Time Series}%\label{sec:timeseries}
In the previous sections, time in the data was aggregated over the whole observation period of four years.
This section is devoted to the temporal development of the network.
Since the trade system changes over time, we first consider the network as a time series of network snapshots.
The temporal resolution of the data set is $\Delta t = 1 d$.
The time series of the network is given by a sequence of $T$ adjacency matrices
\begin{equation}\label{eq:adjmatrixsequence}
\mathcal{A} = \mathbf{A}_1, \mathbf{A}_2, \dots, \mathbf{A}_T,
\end{equation}
where $T$ is the observation period (here: $T=1461\; d$) and each matrix $\mathbf{A}_t$ is the adjacency matrix of the network at time $t$ (snapshot), i.e. it contains the very edges being active at that time.

\begin{figure}[htbp]
\begin{center}
\includegraphics[]{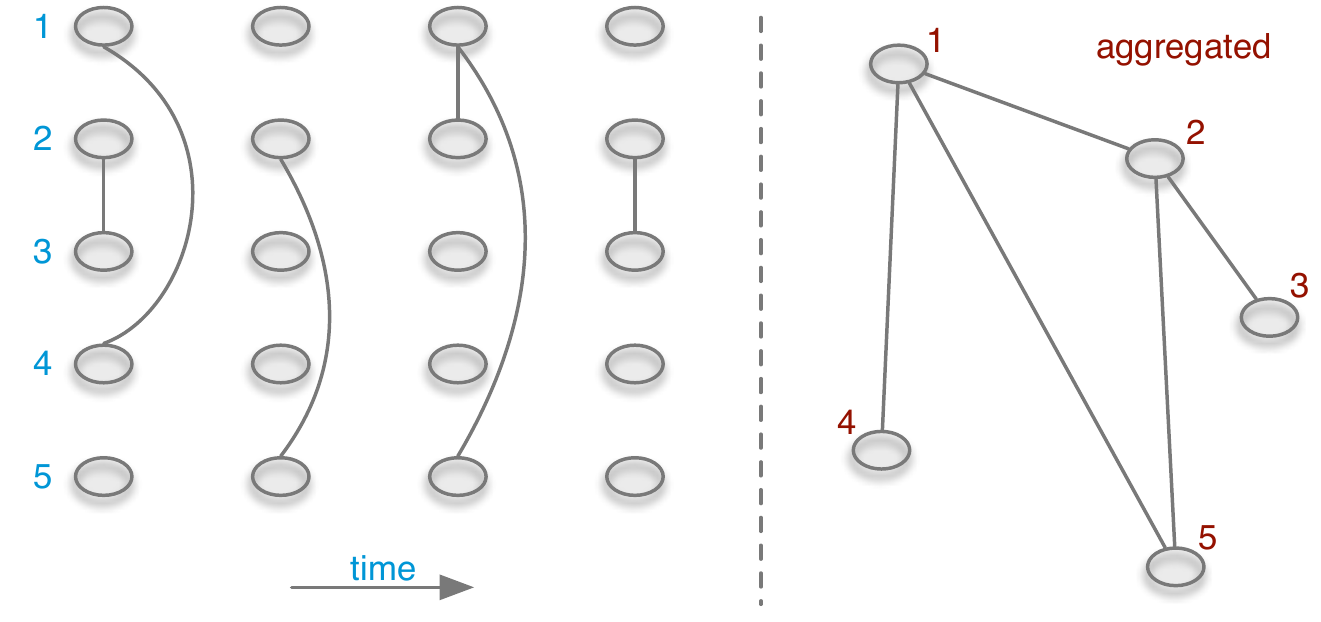}
\caption{A temporal network (left) and its static counterpart (right).}
\label{fig:aggregate_network}
\end{center}
\end{figure}

The static network analyzed above is given by the aggregation over \eqref{eq:adjmatrixsequence}.
Thus the adjacency matrix of the static network is
\begin{equation}\label{eq:aggregation_matrices}
\mathbf{A} = \sum _{t=0} ^T \mathbf{A}_t .
\end{equation}
This directly applies for the aggregation of trade volumes:
for the case of the unweighted network, where volume is not taken into account, the matrices can be treated as Boolean.
This means in effect that every non zero element is set to unity.
Fig.~\ref{fig:aggregate_network} shows the aggregation of an exemplary undirected network.

This raises the question how many snapshots $\mathbf{A}_t$ have to be aggregated (size of the aggregation window) in order to recover the properties of the static network as discussed above.
The minimum aggregation window has been analyzed for a similar data set (supporting information in \cite{Konschake:2013js}).
It is roughly one year.

\begin{figure}[htbp]
\begin{center}
\includegraphics[]{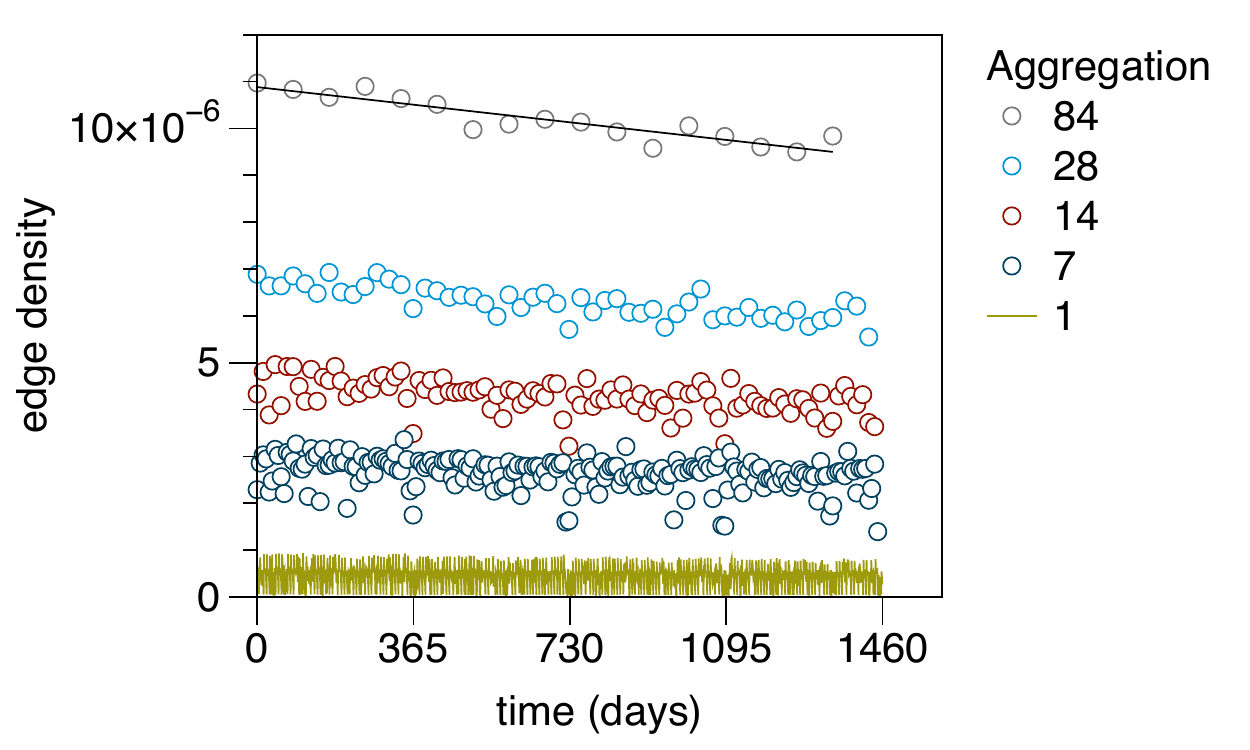}
\caption{Development of the edge density. There is a clear trend to edge reduction over time. The edge density of the static network is $3\times 10^{-5}$.}
\label{fig:edge_act}
\end{center}
\end{figure}

In order to reveal general trends in the temporal evolution of the system, snapshots of the network can be compared at different times.
The temporal evolution of the number of edges is shown in Fig.~\ref{fig:edge_act} (the figure shows the edge density, i.e. the number of edges normalized by the number of theoretical possible edges).
To reduce noise in the data, snapshots have been partially aggregated.
The following partial aggregations have been considered: 1 d (yellow), 7 d (weekly, dark blue), 14 d (red), 28 d (monthly, light blue), 84 d (quarterly, grey).
For an aggregation window of 84 days we find a linear slope of $10^{-9} d^{-1}$; this corresponds to a decrease of about 3,600 edges per year.
The number of active nodes shows a similar trend (\nameref{S5}).
A decrease in the number of active nodes implies that gradually less premises will play a role for disease spread.

Concerning the difference between the static network and the time series, Fig.~\ref{fig:edge_act} shows that the edge density of each snapshot, represented by a matrix $\mathbf{A}_t$, is on average less than $10^{-6}$.
On the other hand, the static network has an edge density of $ 3\times 10^{-5}$ (Table~\ref{tab:properties}).
Hence the edge density of the aggregated network is about an order of magnitude higher, i.e. about 10 \% of the edges are active every day.

\begin{figure}[htbp]
\begin{center}
\includegraphics[]{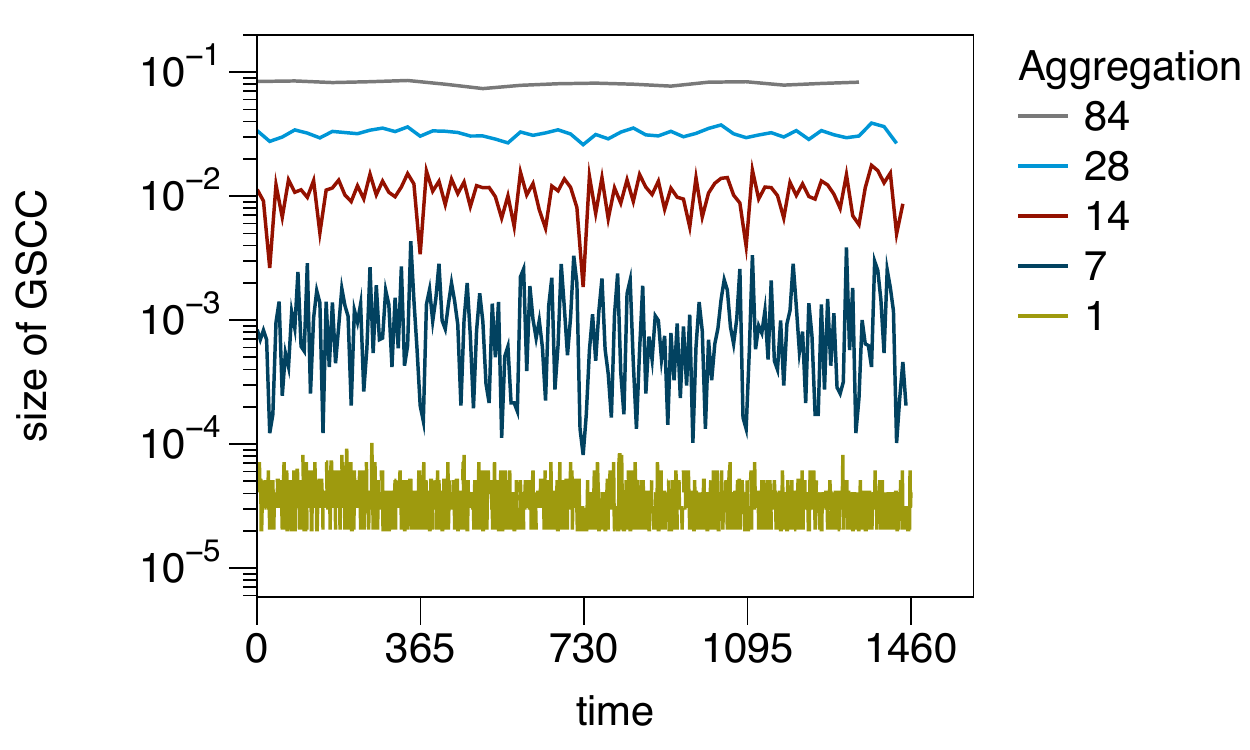}
\caption{Relative size of the GSCC for different partial aggregation windows. The sizes show stronger seasonal fluctuations on small time scales (red), but remain rather constant over large time scales (grey). Size is normalized by the number of nodes in $G$.}
\label{fig:GSCC_act}
\end{center}
\end{figure}

It should be noted that the size of the GSCC is almost unaffected by the trend observed in Fig~\ref{fig:edge_act}.
This is shown in Fig.~\ref{fig:GSCC_act}.
The GSCC shows seasonal fluctuations for intermediate aggregation windows (see also \cite{Kao:2006iz}).
The relatively high stability of the GSCC over time reflects the fact that the network maintains its functionality even though the number of links is reduced over time. 

\begin{figure}[htbp]
\begin{center}
\includegraphics[]{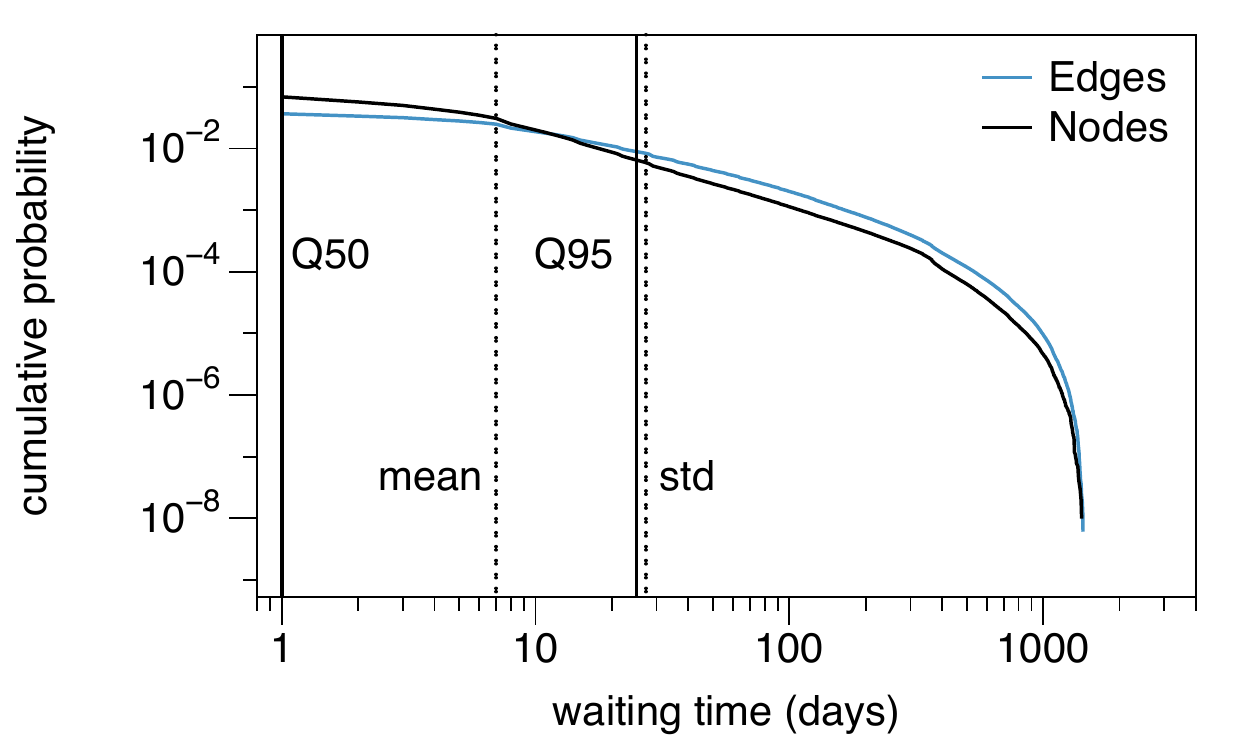}
\caption{Distribution of node and edge waiting times. The empirical waiting times cover values over three orders of magnitude. Dashed lines show mean and standard deviation of the node waiting times, respectively. Solid lines show median and 95 \% quantile of the node waiting time distribution.}
\label{fig:waiting_times_pdf}
\end{center}
\end{figure}

It is important to note that the waiting times in the network are strongly heterogeneously distributed.
The waiting times of a node (or edge) are times where the node (or edge) is not active.
Fig.~\ref{fig:waiting_times_pdf} shows the distribution of the waiting times of nodes and edges.
The figure illustrates that the measured inactivity time spread over several orders of magnitude.
Given the shape of the distribution no appropriate mean value can be given for the waiting times.
%%%% ##### (Knoten: mean = 7.0, std = 27.2; Q05, Q50, Q95 = 0, 1, 25. Kanten: mean = 27.3, std = 62.0; Q05, Q50, Q95 = 1, 10, 107).
This is also an indication that an interpretation of the trade links as \emph{rates} between nodes (e.g. the flux of animals between node $i$ and $j$ is $m$ animals per day) is not appropriate for this system.
A similar behavior has also been found for other system and is referred to as \emph{bursty behavior} \cite{Pan:2011dga, Holme_review, Konschake:2013js, Bajardi:2011iv}.
It should be mentioned that for the pig trade network considered here, typical waiting times might be in the system, if premise types were resolved in the data.
However, this is not the case for the considered data set and thus the inactivity time reflect a global behavior over all nodes.

In conclusion, nodes and edges can be inactive over a long period of time.
This raises the question whether the network can be treated as a static system at all.
After all, edges are in fact considered as permanent in the static network.
We will address this question in the next section.

\subsection*{Temporal Network Analysis}%\label{sec:temporal}
In all methods described above the network was either considered as an aggregated static system or as independent snapshots.
However, a closer look reveals that each snapshot is essentially not a meaningful network for disease spread, since typically \emph{indirect} trade connections are not traversed at a single time step.
For a realistic network traversal, edges at different time steps are necessary.

Thus the network under consideration is in fact a \emph{temporal network}.
Existing results and methods from classical social network analysis cannot necessarily be transferred to temporal networks.
Overviews of temporal network analysis are provided in \cite{Casteights_review, Holme_review, Holme:2015il} and Chapter 4 in \cite{lentz:thesis2013}.
Analyses of epidemic spreading in temporal networks can be found in \cite{Rocha_plosbc,Valdano:2015il,Bajardi:2011iv,Konschake:2013js,Bajardi:2012,Vernon:2009p5068}.

In this article, we choose a fundamental approach to analyze the pig trade network as a temporal network:
the common ground between static networks and temporal networks is the accessibility matrix, i.e. the information whether a node can reach another node via an indirect connection.
These connections are called paths.

Let us first consider a static network $G=(V,E)$.
A path $P_{i \rightarrow j}$ from node $i$ to node $j$ is formally given by a sequence of edges between these nodes where the edges can traverse arbitrary other nodes $x_k \in V$, i.e.
\begin{equation}\label{eq:static_path_def}
P_{i \rightarrow j}= \left[ (i, x_1), (x_1, x_2),\dots, (x_n, j) \right].
\end{equation}
The number of steps is the path length.
In general, a large number of paths exists between each node pair in a network.
For the initial spread of infectious diseases from node $i$ to node $j$, only the \emph{shortest path} is of importance, since any longer path between $i$ and $j$ would just correspond to a repeated infection.

In order to take the dynamic nature of trade (in particular heterogeneously distributed waiting times, Fig.~\ref{fig:waiting_times_pdf}) into account, every edge of the network has to be tagged with a timestamp.
A temporal network is formally given by $\mathcal{G}=(V, \mathcal{E})$, where $V$ is a set of nodes and $\mathcal{E}$ is a set of temporal edges \cite{Casteights_review}.
An edge $(i,j,t) \in \mathcal{E}$ connects nodes $i$ and $j$ at time $t$.
Concerning the static paths defined in Equation~\eqref{eq:static_path_def}, an important difference in temporal networks is the fact that successive edges require timestamps that are successive as well.
In other words a path in a temporal network has to be \emph{causal}.
We refer to a \emph{causal path} from node $i$ to $j$ as $P_{i \rightsquigarrow j}$.
Thus, it follows by analogy to \eqref{eq:static_path_def}
\begin{equation}\label{eq:temp_path_def}
P_{i \rightsquigarrow j}= \left[ (i, x_1, t_1), (x_1, x_2, t_2),\dots, (x_n, j, t_n) \right]
\end{equation}
with the causality constraint
\begin{equation}\label{eq:time_constraint}
t_1 < t_2 < \cdots < t_n .
\end{equation}
The \emph{path duration} is defined as $t_n$.
Consequently, the shortest path duration from node $i$ to $j$ is that connection where $t_n$ in $P_{i \rightsquigarrow j}$ is minimal.

It is important to emphasize that due to the causality constraint, network traversal cannot be carried over from the static to the temporal case in a straightforward manner.
On the other hand, the concept of \emph{accessibility} holds also for the temporal case.
Therefore, we will use this common ground in order to analyze the pig trade in terms of a temporal network.

The accessibility of a static network can be written as a matrix $\mathbf{P}$ with entries:
\begin{equation}\label{eq:static_path_matrix}
(\mathbf{P})_{ij} =
\begin{cases}
1 & P_{i \rightarrow j}  \text{ exists}\\
0 & \text{else}.
\end{cases}
\end{equation}
The accessibility matrix can be interpreted as the adjacency matrix of the corresponding accessibility graph.
For the temporal case, the accessibility matrix $\mathcal{P}$ has the entries:
\begin{equation}\label{eq:temporal_path_matrix}
(\mathcal{P})_{ij} =
\begin{cases}
1 &  P_{i \rightsquigarrow j} \text{ exists}\\
0 & \text{else}.
\end{cases}
\end{equation}

\begin{figure}[htbp]
\begin{center}
\includegraphics{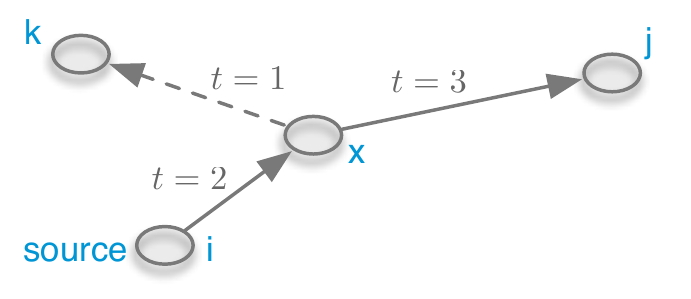}
\caption{Causal path between nodes $i$ and $j$ in a temporal network. Although the path $P _{i \rightsquigarrow k}$ does not exist in the temporal network, this path exists in the static case.}
\label{fig:causal_path}
\end{center}
\end{figure}

Fig.~\ref{fig:causal_path} shows an exemplary causal path between nodes $i$ and $j$.
Although there is no causal path from $i$ to $k$, this path would exist in the static view on the network.
If $i$ was the source node of an epidemic outbreak, the epidemic could never reach node $k$ and a static view on the network would overestimate the outbreak size.

The authors would like to stress the fact that even in the temporal case the accessibility matrix represents a mathematical graph and is a static quantity.
Thus, all concepts above can be transmitted one-to-one from $\mathbf{P}$ (static network) to $\mathcal{P}$ (temporal network).

%\paragraph{Temporal network analysis using paths.}
%In this chapter we describe how the accessibility of a temporal network can be computed using standard mathematical tools.
 
\paragraph{Computation of the Accessibility Matrix.}
Given a \emph{static} network with $N$ nodes the accessibility matrix can be computed as follows \cite{Warshall:1962wr}:
\begin{equation}
\mathbf{P} \sim \sum _n ^N \mathbf{A}^n ,
\end{equation}
where $\mathbf{A}$ is the adjacency matrix of the network.
Nevertheless, more efficient methods can be used here \cite{Skiena:2008:ADM:1410219}.

Given a \emph{temporal} network of $T$ time steps, the accessibility matrix as it was formally defined in \eqref{eq:temporal_path_matrix} can be computed explicitly as follows (\cite{Grindrod:2011fg, Lentz:2013PRL} and Chapter 4.3 in \cite{lentz:thesis2013}):
\begin{equation}\label{eq:tem_access_1} 
\mathcal{P} = \prod _{t=1} ^T (\mathbf{A}_t + \mathbf{1}),
\end{equation}
where $\mathbf{A}_t$ is a snapshot of the network at time $t$ (see Equation~\eqref{eq:adjmatrixsequence}) and $\mathbf{1}$ is the identity matrix.
The accessibility defined in Equation~\eqref{eq:tem_access_1} takes the causality of paths into account.
Following Equation~$\mathcal{P} $, the entries of the accessibility matrix represent the number of paths between node pairs.
In most cases this number is not relevant.
Thus, $\mathcal{P} $ can be treated as a Boolean matrix for convenience, i.e. all non zero elements are set to unity.

\paragraph{Range.}
As we have seen above, the accessibility matrix contains the information whether an infection started at some node $i$ can reach another node $j$ at all.
This has been used implicitly already in Fig.~\ref{fig:ranges_static}, whereby the range of a node $i$ can be computed as $ r_i = \sum _j (\mathbf{P})_{ij}$.

This definition can be transferred to the temporal network case in a straightforward manner.
Once the accessibility matrix is computed, the (temporal) range of a node is
\begin{equation}\label{eq:temp_range}
r_i = \sum _j (\mathcal{P})_{ij} .
\end{equation}
In case of a disease outbreak starting at node $i$, this quantity gives the number of nodes that can potentially be infected.

\begin{figure}[htbp]
\begin{center}
\includegraphics[]{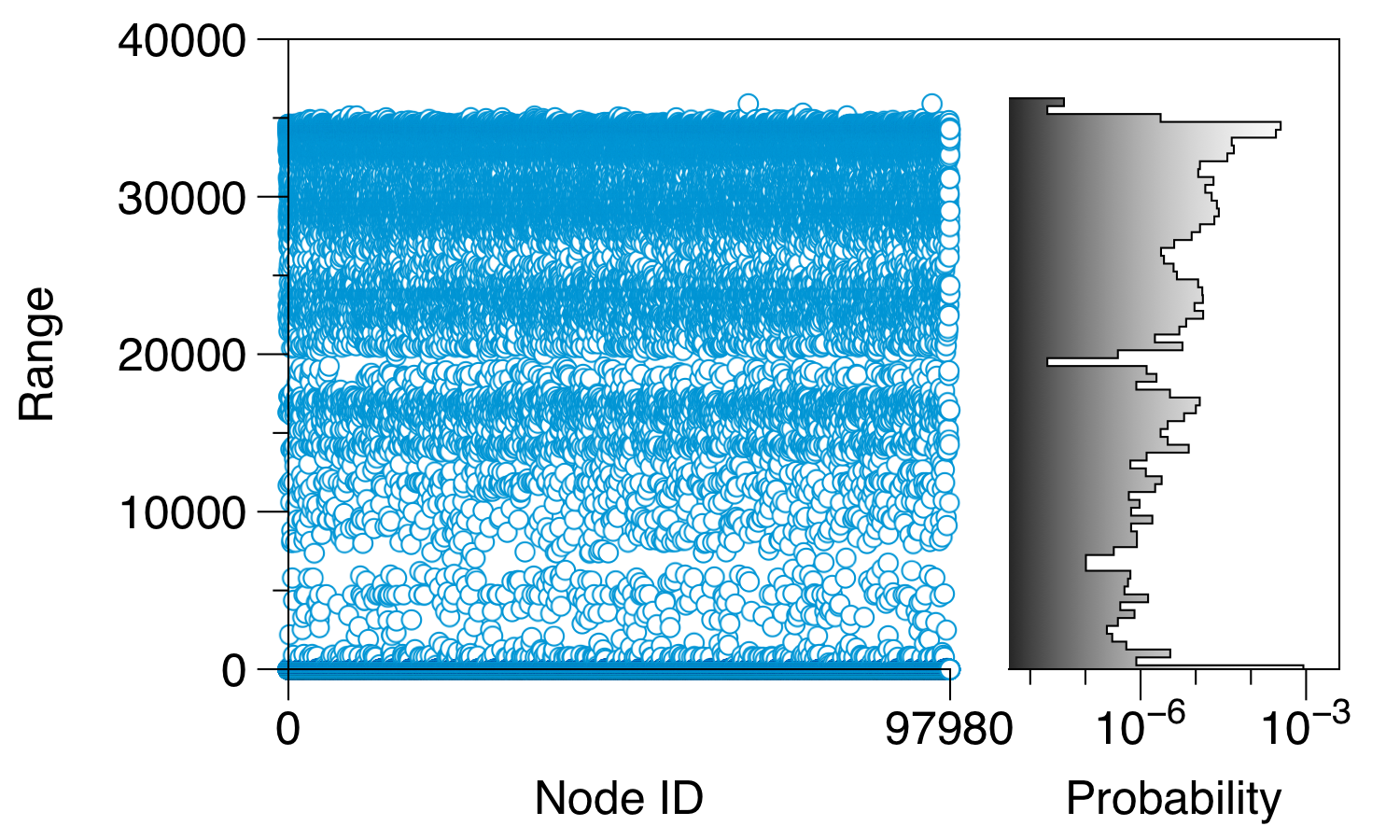}
\caption{Range for every node in the temporal network. The right panel shows the histogram of the y-axis values on a log scale.
In contrast to the static case (see Figure \ref{fig:ranges_static}) the values cover the whole spectrum from minimum to maximum range.}
\label{fig:ranges_temp}
\end{center}
\end{figure}

Fig.~\ref{fig:ranges_temp} shows the range for each node in the temporal pig trade network.
The chart is the counterpart to Fig.~\ref{fig:ranges_static}.
The bimodal distribution as observed in Fig.~\ref{fig:ranges_static} is preserved also for the temporal case, although the shape is less pronounced.
In contrast to the static case, temporal ranges are observed over the whole spectrum of possible values.
This finding suggests that the temporal network does not contain a clear GSCC.
In fact, the concept of connectedness in temporal networks is associated with some conceptional problems \cite{Casteights_review, Nicosia:2012hz}.

The maximum range in the temporal network is 35,905 (for comparison: 41,369 in the static case).
On average the temporal range is 17,186.8 (static: 23,154.2).
In summary, the average size of an outbreak is overestimated in the static case by almost 35~\% and the maximum outbreak size by about 15~\%.
It turns out here already that the analysis of this temporal network gives significantly different outbreak patterns than those observed for the static network representation.
We will define the error of the static representation of a temporal network and implications for epidemiology below.

\paragraph{Path Density.}
The number of edges in the accessibility matrix contains important information about a network.
In case of the adjacency matrix the number of nonzero elements is up to a constant the edge density of the network\footnote{, i.e. $\rho (\mathbf{A}) = (\sum _{ij} (\mathbf{A})_{ij})/N(N-1)$.}.
Analogously, the \emph{path density} of a (static) network is given by $\rho (\mathbf{P}) = (\sum _{ij} (\mathbf{P})_{ij})/N^2$.
The factor $N^2$ is chosen since nodes can have a path back to themselves.

For the temporal case, we define the path density:
\begin{equation} \label{eq:path_density_def}
\rho (\mathcal{P}) = \frac{\sum _{ij} (\mathcal{P})_{ij}}{N^2}.
\end{equation}
The path density takes values between 0 and 1.
It should be noted that Equation~\eqref{eq:path_density_def} holds for Boolean matrices\footnote{In general $ \rho (\mathcal{P}) = \mathrm{nnz} (\mathcal{P})/N^2$, where $\mathrm{nnz} (\mathcal{P})$ is the number of non zero elements of $\mathcal{P}$}.
It contains the information whether a network contains structural holes:
In the limit of a high path density, i.e. $\rho (\mathcal{P}) \approx 1$, most nodes can reach each other.
On the contrary, for a low path density ($\rho (\mathcal{P}) \approx 0$) the network tends to be temporarily disconnected \cite{Nicosia:2012hz}.
For the pig trade network we measure $\rho (\mathbf{P})\approx 0.24$ for the static case (see table \ref{tab:properties}) and $\rho (\mathcal{P}) \approx 0.18$ for the temporal case.

\paragraph{Comparison between Static and Temporal Network Representation.}
The static network as described above is an approximation of the temporal system.
This approximation is obtained by temporal aggregation and this means a removal of causality in paths.
As stated above, causality plays an important role for the traversal of temporal networks.
This raises the question to what extend a static network representation reflects the real causal accessibility between node pairs correctly.

\begin{figure}[htbp]
\begin{center}
\includegraphics[]{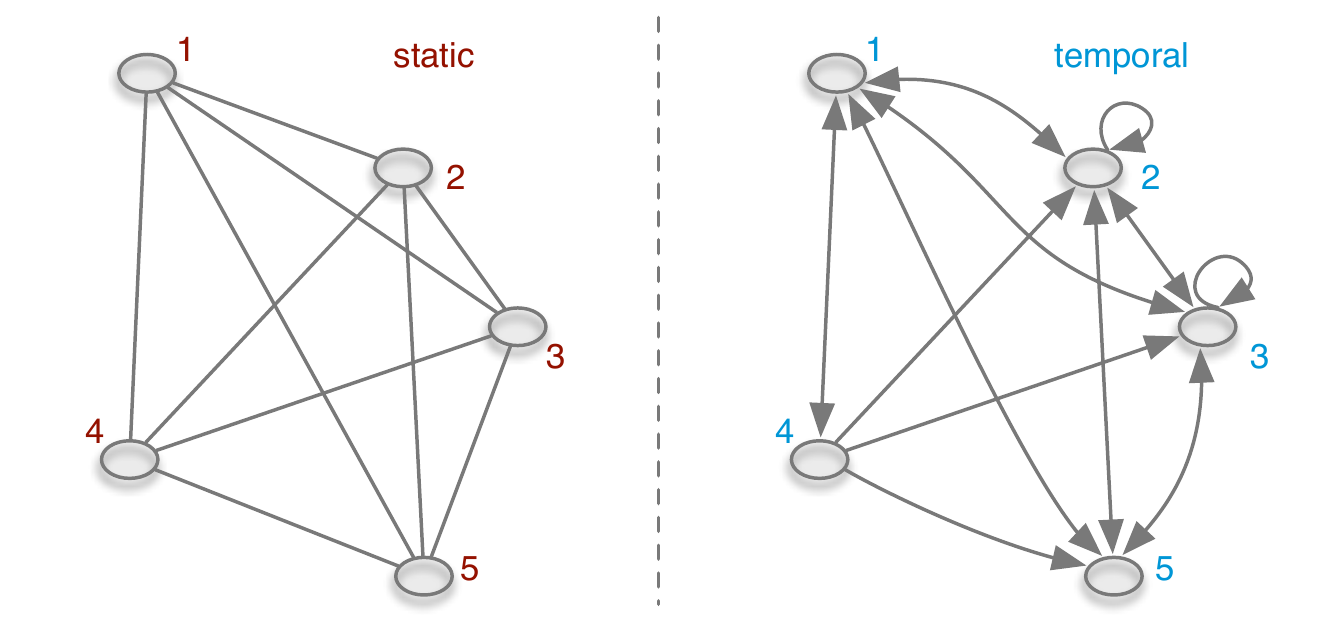}
\caption{Accessibility graphs of the network of Figure \ref{fig:aggregate_network}. All nodes in the static accessibility graph (left) have a path back to themselves (i.e. self-loops, not shown).
Note that although the underlying temporal network is undirected, the temporal accessibility graph is directed.}
\label{fig:causal_principle_example}
\end{center}
\end{figure}

The difference between the accessibility of a temporal network and its static representation is illustrated in Fig.~\ref{fig:causal_principle_example}.
If the network is aggregated over time, a path from every node to every other node in the network would be present, i.e. it exists $P_{i \rightarrow j}$ for all nodes $i$ and $j$ including paths from a node back to itself (so-called \emph{self-loops}).
For the temporal case the following paths do not exist: $P_{2\rightsquigarrow 4}$, $P_{3 \rightsquigarrow 4}$, $P_{5 \rightsquigarrow 4}$ and the self-loops $P_{1 \rightsquigarrow 1}$, $P_{4 \rightsquigarrow 4}$ and $P_{5 \rightsquigarrow 5}$.
It should be noted that the consideration of self-loops is a matter of definition.
In large systems (as the pig trade network) self-loops are statistically irrelevant, i.e. the number of possible self-loops is small (order of $N$) compared to the number of possible paths (order of $N^2$).
In addition, Fig.~\ref{fig:causal_principle_example} demonstrates an interesting feature about accessibility graphs of temporal networks:
even if the underlying network is undirected, the accessibility graph of a temporal network is generally directed.

In order to quantify the error of the static representation of a temporal network, the number of paths in the static view can be compared to the number of paths in the temporal system \cite{Lentz:2013PRL}.
Their ratio is called \emph{causal fidelity} $c$, where
\begin{equation}\label{eq:causal_fid_def}
c = \frac{\sum _{ij} \mathcal{P}_{ij}}{\sum _{ij} \mathbf{P}_{ij}} =  \frac{\rho(\mathcal{P})}{\rho( \mathbf{P})} =  \frac{\text{number of paths in } \mathcal{G}}{\text{number of paths in }\mathbf{G}}.
\end{equation}
The number of paths is the number of non zero elements in $\mathcal{P}$ or $\mathbf{P}$, respectively.
A causal fidelity of $1$ means that a temporal network is well represented by its static counterpart.
On the other hand, when $c \approx 0$, the network should not be considered as a static system, since most paths are not causal.

In the example in Fig.~\ref{fig:causal_principle_example}, there are mutual paths between all five nodes (including self-loops), i.e. $\sum _{ij} \mathbf{P}_{ij} = 5^2 =25$.
On the other hand, in the temporal case we have $\sum _{ij} \mathcal{P}_{ij} = 19$ paths.
Thus, the causal fidelity is $c=19/25$.

For the pig trade network, we measure a causal fidelity of
\begin{equation}\label{eq:causal_fid_D}
c _\text{pig trade} = \frac{1{,}683{,}966{,}477}{2{,}268{,}652{,}889} \approx 0.74 .
\end{equation}
This implies that 26 \% of the paths that appear to be present in the static network, do not actually exist.
As already indicated above, the reciprocal causal fidelity gives an estimation, to what extend a static network view would overestimate a disease outbreak.
Therefore, we define the \emph{causal error} of the static network as
\begin{equation}\label{eq:causal_error}
\epsilon _\text{pig trade}=  \frac{1}{c _\text{pig trade}}  \approx 1.35 .
\end{equation}
This value refers to the number of potentially infected nodes for a worst case outbreak scenario over the whole observation time.
Consequently, in such a scenario a static representation of the pig trade network would overestimate the size of a disease outbreak by a factor of $1.35$.
It should be mentioned that the causal error is not normalized as this is the case for causal fidelity.

\paragraph{Unfolding Accessibility.}
In the previous section, we discussed how an accessibility graph can be computed.
We hereby took into account the whole available time period.
Nevertheless there is more information in the accessibility graph.
In short, this information can be retrieved if the accessibility matrix is computed step by step and the path density $\rho $ (see \eqref{eq:path_density_def}) is stored at every step.
Hereby, we want to answer the following questions:
\begin{enumerate}
\item how can the dynamics of an outbreak be modeled in a temporal network?
\item what is the expected time scale of such an outbreak?
\end{enumerate}
The second question aims at the fact that a temporal network exhibits not only a topological path length, but also a path \emph{duration} (see Equation~\eqref{eq:time_constraint}).
It is indeed possible that the average shortest path length of a network is short compared to the network size (see Table~\ref{tab:properties}), but the path duration is very long.
In other words, even a short path can take a lot of time.
This information is of major interest for disease control since it provides an estimation of the time scale of a disease.

In order to answer the questions above, we consider the accessibility matrix as defined in Equation~\eqref{eq:tem_access_1}, but we consider $T$ in Equation~\eqref{eq:tem_access_1} as the evolving time $t<T$.
Hereby, we stepwise store the current result at time $t$, i.e.
\begin{equation} \label{eq:unfolding_acc}
\mathcal{P}(t) = \prod _{t'=1} ^t (\mathbf{A}_{t'} + \mathbf{1}) .
\end{equation}
Equation~\eqref{eq:unfolding_acc} yields the temporal evolution of the accessibility.
The process of the stepwise computation is referred to as \emph{unfolding accessibility} \cite{Lentz:2013PRL}.
Hereby, the focus is on the path \emph{density} since it is a real number (and not a matrix).

Starting at $t=1$, the matrix $\mathcal{P}(t)$ contains self-loops and the paths to the nearest neighbors.
The former is a necessary artifact to allow for paths after inactive periods (see~\cite{Lentz:2013PRL} for details).
At $t=2$ the matrix contains all new paths at that time step as well as all former paths and so forth.
Thus, the path density grows with every time step.
This process mimics an SI-type (susceptible -- infected) spreading process with infection probability 1 on the network.
That means every causal path is a potential route along which a disease can spread and a node is exposed, whenever it lies on such an infective path.

\begin{figure}[htbp]
\begin{center}
\includegraphics[]{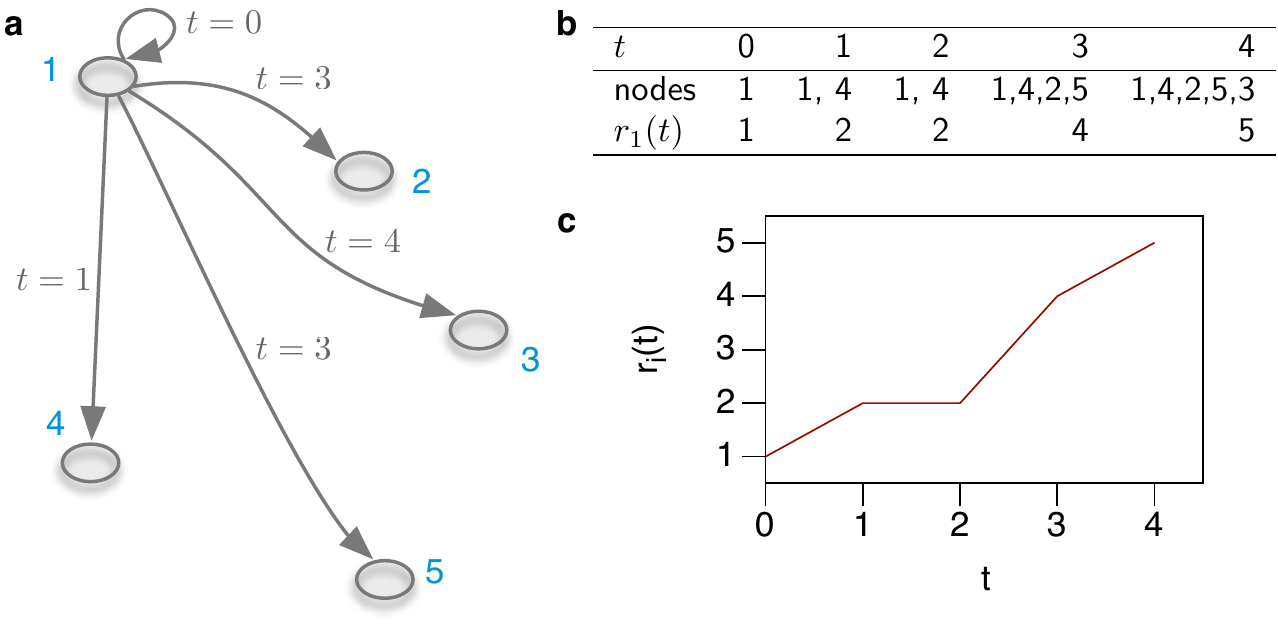}
\caption{Unfolding accessibility for node 1 in the network shown in Fig.~\ref{fig:aggregate_network}. \textbf{a} Accessibility of node 1 including time stamps when nodes are accessed. \textbf{b} Number of infected nodes over time. \textbf{c} Infection curve (i.e. range) for source node 1.}
\label{fig:unfolding_single_node_ex}
\end{center}
\end{figure}

In analogy to the range defined above \eqref{eq:temp_range}, the current range $r_i(t)$ of a node $i$ is given by
\begin{equation}\label{eq:outbreak_size_range}
r_i (t) = \sum _j ( \mathcal{P} (t) )_{ij} .
\end{equation}
The \emph{herd prevalence} of an SI-process is then given by $r_i(t)/N$.
As an example, Fig.~\ref{fig:unfolding_single_node_ex} shows the current range of a node in the network from Fig.~\ref{fig:aggregate_network}.

In order to obtain the average herd prevalence $\bar{r}(t)/N$, one can average over all starting nodes, i.e.
\begin{equation}\label{eq:outbreak_size_total}
\bar{r}(t) = \sum _{ij} ( \mathcal{P} (t) )_{ij} = \rho (\mathcal{P}(t)) .
\end{equation}
Using Equations \eqref{eq:outbreak_size_range} and \eqref{eq:outbreak_size_total} the first question is already answered.
In short, an SI-process can be modeled by calculating the temporal evolution of the accessibility matrix.
Hence, the path density at every time step corresponds to the average herd prevalence over all starting nodes.

In order to answer the second question, we have to find the distribution of path durations.
Considering again the new established paths at every time step, $\rho (\mathcal{P}(t=2))$ for example contains the number of new paths at time $t=2$ \emph{plus} the number of paths at $t=1$ and so forth.
In fact, this corresponds to the cumulative distribution of shortest path durations.
Consequently, $F_t = \rho (\mathcal{P}(t))/N^2$ is the cumulative distribution function (CDF) of path durations in a temporal network\footnote{In this definition the cumulative distribution function is not necessarily normalized. We consider them as 'improper' distribution functions.}.
The desired shortest path duration distribution is given by $dF_t/dt$.

\begin{figure}[htbp]
\begin{center}
\includegraphics[]{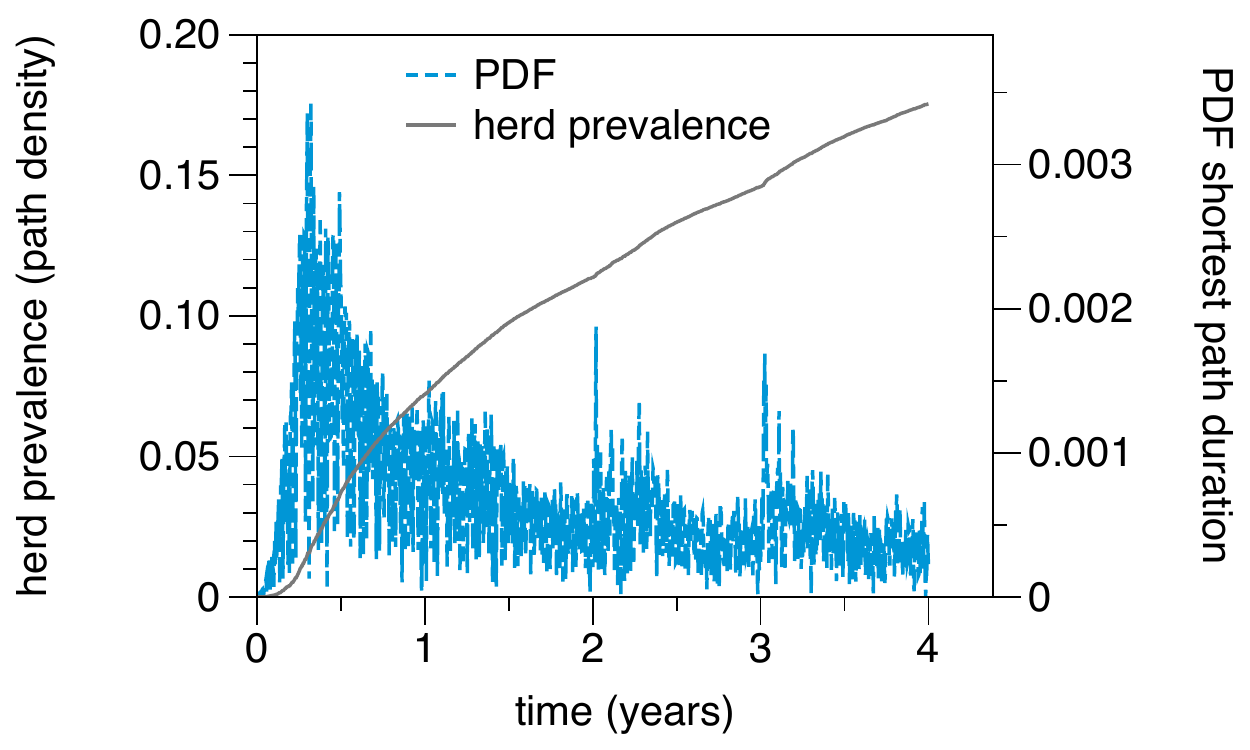}
\caption{Unfolding accessibility of the temporal pig trade network. The mean herd prevalence (grey solid line) is given by the path density. The probability density function (PDF) of shortest path durations (blue dotted line) shows a global maximum at about 120 days.}
\label{fig:unfolding}
\end{center}
\end{figure}

Fig.~\ref{fig:unfolding} shows the path density (grey solid line) and the probability distribution of shortest path durations (blue dashed line) of the network \cite{Lentz:2013PRL}.
The path density corresponds to the mean infection curve.
It shows the typical shape of an SI-infection curve, although no saturation is reached due to limited observation time in the data set.
The shortest path duration (blue dotted line) shows a significant maximum at around half a year.
It reaches its peak at about 120 days.
%This means that an infection spread by pig trade in Germany takes on average 120 days to spread over the network.
This means that the majority of paths for infection spread by pig trade in Germany take 120 days.
% This means that for pig trade in Germany the majority of paths take 120 days for infection spread.
Roughly speaking, 120 days is a typical time scale for infection spread.
It is important to stress the fact that this time scale does not depend on any specific disease parameters, but is a pure property of the network as a substrate for spreading.
An explanation for this time scale can be found in the structure of the underlying pork production chain (see Figure \ref{fig:pork_prod_chain}).
It defines the temporal diameter of the network (180 days).
As observed in Fig.~\ref{fig:unfolding}, this diameter should limit the distribution of shortest path durations.
The maximum is below that value, since there are more possible shorter paths in the production chain than the maximum (longest) path.

\paragraph{Causal Contact Tracing.}
The unfolding accessibility method explained above can be used in a straightforward manner for contact tracing in temporal networks.
As an addition to existing contact tracing software \cite{remark:2014hw}, the method proposed here provides a contact tracing where concepts such as causal error and path density can be analyzed mathematically.

Tracing forward over a certain period $\tau $ is equivalent to an accessibility unfolding from the assumed date of entry to $t=\tau $.
Using our method, possible paths for infection are computed for all source nodes at once.
However, if one is only interested in contact premises of one single index premise, the equation for unfolding accessibility \eqref{eq:unfolding_acc} can be rewritten accordingly.
Let the infection status of the network at time $t$ be given by a row-vector $\mathbf{x}(t)$ with $x_i = 0$, where $x_i = 0$ if node $i$ is susceptible and $x_i \neq 0$ if node $i$ is infected.
If node $j$ is the only node infected at the starting time, then the initial vector is $\mathbf{x}(0)$ where $x_j = 1$ and $x_i=0$ otherwise.
The newly infected nodes for every time step are given by the vector
\begin{equation} \label{eq:tracing_fwd}
\mathbf{i}(t+1) = \mathbf{x}(t+1)- \mathbf{x}(t) = \mathbf{x}(t) \mathbf{A}_t.
\end{equation}
This equation follows immediately from Equation~\eqref{eq:unfolding_acc}
The sequence $\mathbf{i}(0), \mathbf{i}(1), \dots , \mathbf{i}(t)$ represents the causal tree of possible contacts of the index node up to time $t$.

In case of a disease outbreak one is also interested in tracing backward.
Therefore, the network has to be traversed backwards in time.
Here we can again make use of Equation~\eqref{eq:tracing_fwd}, but the network has to be time reversed in the first place.
If a temporal network is given by a sequence of adjacency matrices \eqref{eq:adjmatrixsequence}, then the time reversed network is
\begin{equation} \label{eq:time_rev}
\mathcal{A}^{-1} =  \mathbf{A} ^T _T, \dots , \mathbf{A}_2 ^T, \mathbf{A}_1 ^T,
\end{equation}
where $\mathbf{A}_i ^T$ is the transposed of the $i$-th matrix in the sequence.
In other words, the edge direction in each snapshot as well as the temporal ordering of the matrices is reversed.
In order to realize a tracing backward,  the new adjacency matrix sequence \eqref{eq:time_rev} can be used in \eqref{eq:tracing_fwd}.

Depending on the context, it might be reasonable to allow the traversal to have multiple edges within a single time step.
This is the case if there are causal contact chains below the temporal resolution of the network (here $1d$).
As an example, a premise could buy animals from farm $i$ in the morning and sell animals to farm $j$ in the afternoon.
The path from $i$ to $j$ would not be considered using the approach above.
Another example is bad reporting compliance in the sense that multiple transactions might be reported for the same day, but actually happened at multiple points in time.
If such circumstances have to be taken into account, we call the tracing procedure \emph{prudent contact tracing}.

In this case, longer static paths have to be considered for every snapshot of the system \cite{Grindrod:2011fg}.
Therefore, outbreaks are larger in general.
For single nodes, this can have a considerable impact on the number of possible contact nodes.
Nevertheless, if the network is considered as a whole, the effect is rather small and results do not change qualitatively (\nameref{S6}).

\paragraph{Can we trust the static network representation?}
The found results raise the question whether a static network should be used at all as a substrate for the spread of an infectious disease.
The path structure and the causal fidelity have demonstrated that for the reason of causality alone there is a discrepancy between both views.
In addition, it is important to stress the fact that the concept of time does not per se exist in static networks.
Due to the small average shortest path length in the static network (small world effect, Table~\ref{tab:properties}), simple network traversal models of disease spread would result in unrealistic time scales.
Therefore, any outbreak model on the static network requires the definition of some dynamic process, which includes the definition of parameters.
The time scale of such a process might, among other things, be influenced by the network topology — for instance speed-up by degree correlations (see Section \emph{Mixing Patterns}) — but waiting times on the nodes are not considered.
However, these waiting times play a central role in the form of production times particularly for production networks, such as the pig trade network considered here.
They substantially define the time scale for network traversal.

Furthermore, it should be noted that some measures defined for static networks might be defined in a more complex way (or even not at all) for the temporal case.
As an example the shortest path distance in static networks has three different counterparts in temporal networks \cite{Casteights_review,lentz:thesis2013}.

Nevertheless, the static network model is certainly not redundant due to these circumstances.
For many applications in veterinary medicine, centrality measures are of great importance.
On static networks these measures can be easily defined, computed and interpreted.
Furthermore, it has been shown that some static measures show a good correlation with those for the temporal case \cite{Konschake:2013js}.
Hence, centrality measures computed for the static representation remain relevant also for the temporal case.
In the context of risk based interventions, results from a temporal network analysis could be used in order to improve the quality of static centrality estimations.

\begin{figure}[htbp]
\begin{center}
\includegraphics[]{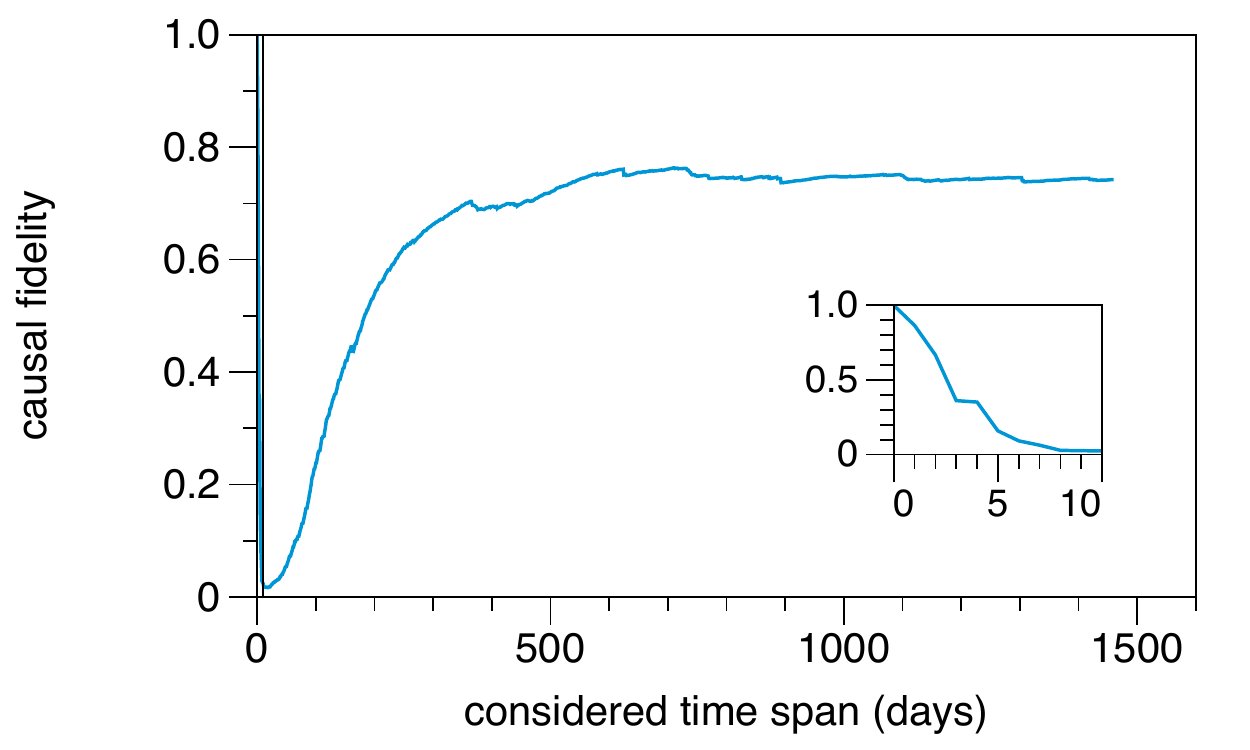}
\caption{Causal fidelity for static networks of different aggregation windows. For aggregation windows $<$ 365 days, the static network representation should be used with caution. A static network view is also adequate for very short aggregation windows (inset).}
\label{fig:cf_unfolding}
\end{center}
\end{figure}

Finally, we focus on the optimum aggregation window for a temporal network, such that the aggregated, static network captures causality sufficiently.
Whether a network can be considered as a static one is determined by the causal fidelity in the first place.
Strictly speaking the causal fidelity depends on the considered time span.
This time span corresponds to the aggregation window used for the static network view.
Fig.~\ref{fig:cf_unfolding} shows the causal fidelity for different aggregation windows.
That is, every day $x$ on the $x$-axis means that the dataset is considered from 01/01/2011 until 01/01/2011 + $x$ days.
For very short time scales (here up to 2 days) the connectivity between node pairs is provided by single edges, i.e. paths of length 1.
Since causality is always maintained for paths of this length, the statistical chance for a break in the causal chain is low.
Therefore, the static view performs well in this range.
However, it should be noted that such a small aggregation window almost corresponds to the fully time resolved temporal network.

For intermediate time scales (2-180 days) already longer paths appear, whereby only a small number of paths exist between each node pair.
This explains the low causal fidelity in this range.
Between 180 and 365 days, the number of paths between each node pair increases strongly until a relatively constant causal fidelity is found for more than 365 days.

In conclusion, the static network representation provides a good picture of the real topology for very small aggregation windows ($< $ 3 days) and for aggregation windows $>$ 365 days.
In the intermediate range, results drawn from a static network representation should be treated with caution.

\section*{Conclusion}

\subsection*{Summary and Discussion}
In this article, we have analyzed the pig trade network in Germany with respect to its ability to spread infectious diseases.
We thereby put a focus on an analysis free of parameters.
Thus, the obtained results do not depend on specific disease models.
Central questions were:
%(1) what is the big picture of the system and (2) where should efficient disease control measures be located?
(1) what is the large scale structure of the system, (2) where should efficient disease control measures be located and (3) what amount of data should be used in order to obtain an appropriate picture of possible spreading paths?

On a global scale, the directed nature of trade plays a crucial role:
the network exhibits a large scale component structure, which in turn causes a sharp classification of the nodes into two risk classes.
Groups of nodes having a large risk of infecting large parts of the system can be found using the ranges of the nodes.

Besides the component structure, the network has a tendency to form subgroups, where little mixing occurs between these subgroups.
In particular, the federal states of Germany show such a behavior.
This result suggests that it might be possible to establish zones or compartments for the German pig production.
Zoning and compartmentalization are tools to define regions within a country with a certain health status to limit the trade restrictions for diseases \cite{oie:2015_code} (e.g. african swine fever in Lithuania \cite{Euro-Lex2}).
Hence, these zones/compartments might be used as a basis for a contingency plan.

Furthermore, there is a weak tendency for nodes of small degree to connect with nodes of large degree.
This fact can be used to make disease control more effective, since vaccination of large degree nodes provides local `firewalls' for all small degree nodes attached to them.
The structure of the pork production system suggests that a significant contribution to the degree mixing pattern is made by farms and slaughterhouses as well as piglet producers and other farms.
However, the central part of the production chains (without slaughterhouses and piglet producers) shows this property as well.
Hence, we conclude that disease control measures can be efficiently applied in the considered network.
%disassortative degree pattern is particularly formed between farms and slaughterhouses and also between piglet producers and other farms.

The efficiency of disease control measures has additionally been investigated by targeted node removal.
For this purpose, different centrality measures have been computed for the nodes in order to obtain a risk based node ranking.
We found that it is sufficient to remove 0.1 \% of the nodes in order to disassemble the network into small islands.
For comparison, in the case of random node removal, 30 \% of the nodes would have been required to obtain the same result.
%It is necessary for answering the questions above to discuss the depth of the considered data set.
%In this case, this means the trade volume and the temporal resolution of the data.

The results discussed above hardly change when the trade volume (number of traded animals) is taken into account.
We conclude therefore that the relevance of the trade volume is of secondary importance in this context.

The authors would like to emphasize that a static consideration of the network shows significant shortcomings for the understanding of disease dynamics.
To begin with, a static network does not contain any time scale by definition.
Hence, the duration of an epidemic for example cannot be estimated from the analysis of network topology alone.
Considering the network as a series of aggregated snapshots improves the results, but does not take into account causality for infectious trade paths.
In addition to that, possible outbreak sizes are systematically overestimated in a static view on the network.

In order to avoid the shortcomings of the static analysis, we continued to regard the system as a temporal network.
We thereby took into account the causal occurrences of edges explicitly.
Using the unfolding accessibility method, we were able to extract temporal information about a potential epidemic outbreak.
This approach mimics an SI-spreading process on the temporal network, where a worst case scenario, i.e. a transmission probability of one is assumed.
It is therefore also an appropriate tool for causal contact tracing.
Overall, the unfolding accessibility method provides both a large scale view on the network and can be used in order to detect central points in the network.

The most probable time for a disease to reach an arbitrary node in the network is 120 days.
Even though it takes on average only 5.5 steps (see Table \ref{tab:properties}) to traverse the network, these steps take in most cases 120 days.
Thus, the network shows the small world, but also the slow world property.

The error of the static view on the actually temporal system can be quantified using the causal fidelity measure, which compares the possible outbreak sizes in the static and the temporal system, respectively.
The causal error of the static network is approximately $\epsilon =1.35$, i.e. a static view on the network overestimates the maximum possible outbreak size by 35 \%.

Although the static network representation exhibits a causal inconsistency that can be quantified using causal fidelity, a static network still remains a valuable model of the pig trade system.
The causal error of $\epsilon =1.35$ is still justifiable for a number of results and therefore a static view of the pig trade systems reflects the topological structure adequately.
A temporal view should be used in particular for intermediate time scales (up to 365 days).

\subsection*{Outlook}
The analyzed dataset reflects trade contacts between holdings of a production network.
However, the different production types of the single nodes are not resolved in the data.
If these data were available, a risk assessment based on different production types could be provided as it has been done in \cite{Buttner2013418}.
This would significantly improve the classification of holdings into risk classes.
In addition, single production chains (see Fig.~\ref{fig:pork_prod_chain}) could be reconstructed, if production type data were available.
As a result, in case of an outbreak it would be possible to implement trade restrictions, while unaffected production chains remain fully functional.

The results provided here can possibly be transferred to similar production systems, but they are still computed for a particular dataset.
In order to obtain a better understanding of the results found here on a more general level, more generative network \cite{Holme:2015il} and dynamic \cite{Wang20133869} models could be applied.

\section*{Supporting Information}

\subsection*{S1 Assortativity Coefficients}
\label{S1}
{\bf Assortativity Coefficients for the Directed Network.}

\subsection*{S2 Targeted Vaccination}
\label{S2}
{\bf Targeted vaccination for other centrality measures (with Figure S1).} These are: Eigenvector centrality, PageRank and Katz-Centrality.

\subsection*{S3 Centrality in Components}
\label{S3}
{\bf Distribution of the centrality over the components of the network  (with Figure S2).}

\subsection*{S4 Weighted Network}
\label{S4}
{\bf Analysis of the weighted network (with Figures S3-S4).} Weighted mixing patterns, weighted centrality, impact of targeted vaccination and analysis of the infection probability per link.

\subsection*{S5 Node Activity over Time}
\label{S5}
{\bf Number of active nodes over time (with Figure S5).}

\subsection*{S6 Prudent Contact Tracing }
\label{S6}
{\bf Temporal range of the nodes, when multiple steps are allowed in each time step (with Figures S6-S7).} Includes a comparison of the path densities for the normal and the prudent case.

%\textbf{S1 Assortativity coefficients
%\newline S2 Targeted vaccination (with Figure S1)
%\newline S3 Centrality in components (with Figure S2)
%\newline S4 Weighted network (with Figures S3-S4)
%\newline S5 Node activity over time (with Figure S5)
%\newline S6 Prudent contact tracing (with Figures S6-S7)
%}

\section*{Acknowledgements}
AK and PH acknowledge funding by the Deutsche Forschungsgemeinschaft in the framework of Collaborative Research Center 910. PH was partially supported by the Federal Ministry of Education and Research (BMBF), Germany (grant no. 01GQ1001B).

%----------------------------------------------------------------------------------------
%	REFERENCE LIST
%----------------------------------------------------------------------------------------
% \bibliographystyle{plos2015.bst}
% \bibliography{Bibo.bib, Bibo_SI.bib}

\begin{thebibliography}{10}

\bibitem{Buttner2013418}
B{\"u}ttner K, Krieter J, Traulsen A, Traulsen I.
\newblock Static network analysis of a pork supply chain in Northern
  Germany---Characterisation of the potential spread of infectious diseases via
  animal movements.
\newblock Prev Vet Med. 2013;110(3--4):418 -- 428.
\newblock Available from:
  \url{http://www.sciencedirect.com/science/article/pii/S0167587713000287}.

\bibitem{fevre_animal_2006}
F{\`e}vre EM, Bronsvoort BMdC, Hamilton KA, Cleaveland S.
\newblock Animal movements and the spread of infectious diseases.
\newblock Trends Microbiol. 2006 Mar;14(3):125--131.
\newblock Available from:
  \url{http://www.sciencedirect.com/science/article/pii/S0966842X06000175}.

\bibitem{green_modelling_2006}
Green DM, Kiss IZ, Kao RR.
\newblock Modelling the initial spread of foot-and-mouth disease through animal
  movements.
\newblock P Roy Soc Lond B Bio. 2006 Nov;273(1602):2729--2735.
\newblock Available from:
  \url{http://rspb.royalsocietypublishing.org/content/273/1602/2729}.

\bibitem{ribbens_transmission_2004}
Ribbens S, Dewulf J, Koenen F, Laevens H, Kruif Ad.
\newblock Transmission of classical swine fever. {A} review.
\newblock Vet Quart. 2004 Dec;26(4):146--155.
\newblock Available from:
  \url{http://dx.doi.org/10.1080/01652176.2004.9695177}.

\bibitem{thrusfield_veterinary_2007}
Thrusfield M.
\newblock Veterinary epidemiology.
\newblock Wiley-Blackwell; 2007.

\bibitem{Mayer:2008ip}
Mayer D, Reiczigel J, Rubel F.
\newblock A Lagrangian particle model to predict the airborne spread of
  foot-and-mouth disease virus.
\newblock Atmos Environ. 2008;42(3):466 -- 479.
\newblock Available from:
  \url{http://www.sciencedirect.com/science/article/pii/S1352231007008643}.

\bibitem{Ducheyne:2007gp}
Ducheyne E, De~Deken R, B{\'e}cu S, Codina B, Nomikou K, Mangana-Vougiaki O,
  et~al.
\newblock {Quantifying the wind dispersal of Culicoides species in Greece and
  Bulgaria.}
\newblock Geospat Health. 2007 May;1(2):177--189.
\newblock Available from:
  \url{http://geospatialhealth.net/index.php/gh/article/view/266}.

\bibitem{MartinezLopez:2011db}
Mart{\'\i}nez-L{\'o}pez B, Ivorra B, Ramos AM, S{\'a}nchez-Vizca{\'\i}no JM.
\newblock {A novel spatial and stochastic model to evaluate the within- and
  between-farm transmission of classical swine fever virus. I. General concepts
  and description of the model}.
\newblock Vet Microbiol. 2011 Jan;147(3-4):300--309.
\newblock Available from: \url{http://dx.doi.org/10.1016/j.vetmic.2010.07.009}.

\bibitem{Olofsson:2014gk}
Olofsson E, N{\"o}remark M, Lewerin SS.
\newblock {Patterns of between-farm contacts via professionals in Sweden.}
\newblock Acta Vet Scand. 2014;56(1):70--13.
\newblock Available from: \url{http://www.actavetscand.com/content/56/1/70}.

\bibitem{Wang2012543}
Wang Y, Jin Z, Yang Z, Zhang ZK, Zhou T, Sun GQ.
\newblock Global analysis of an SIS model with an infective vector on complex
  networks.
\newblock Nonlinear Anal Real. 2012 01;13(2):543 -- 557.
\newblock Available from:
  \url{http://www.sciencedirect.com/science/article/pii/S1468121811001945}.

\bibitem{oie:2015_code}
OiE.
\newblock Terrestrial Animal Health Code.
\newblock 24th ed. OiE; 2015.

\bibitem{KnightJones2013161}
Knight-Jones TJD, Rushton J.
\newblock The economic impacts of foot and mouth disease -- What are they, how
  big are they and where do they occur?
\newblock Preventive Veterinary Medicine. 2013;112(3--4):161 -- 173.
\newblock Available from:
  \url{http://www.sciencedirect.com/science/article/pii/S0167587713002390}.

\bibitem{Fritzemeier:2000p5909}
Fritzemeier J.
\newblock {Epidemiology of classical swine fever in Germany in the 1990s}.
\newblock Vet Microbiol. 2000 Nov;77(1-2):29--41.
\newblock Available from:
  \url{http://linkinghub.elsevier.com/retrieve/pii/S0378113500002546}.

\bibitem{faostat}
{FAO - Economic and Social Development Department}.
\newblock FAO [website]; 2016 [cited 2016].
\newblock Available from: \url{http://faostat.fao.org/}.

\bibitem{bmel:2015}
Federal~Ministry of~Food; Agriculture (BMEL).
\newblock Understanding Farming - Facts and figures about German farming.
\newblock BMEL; 2015.
\newblock Available from:
  \url{http://www.bmel.de/SharedDocs/Downloads/EN/Publications/UnderstandingFa%
rming.html;jsessionid=45E09F8A934E452B87D762698DBE305C.2_cid296} [cited 2015].

\bibitem{MartinezLopez2009}
Mart{\'\i}nez-L{\'o}pez B, Perez AM, S{\'a}nchez-Vizca{\'\i}no JM.
\newblock {Social Network Analysis. Review of General Concepts and Use in
  Preventive Veterinary Medicine}.
\newblock Transbound Emerg Dis. 2009;56:109--120.

\bibitem{Dube:2009gx}
Dub{\'e} C, Ribble C, Kelton D, Mcnab B.
\newblock {A Review of Network Analysis Terminology and its Application to
  Foot-and-Mouth Disease Modelling and Policy Development}.
\newblock Transbound Emerg Dis. 2009 Apr;56(3):73--85.

\bibitem{Bigras:2007}
Bigras-Poulin M, Barfod K, Mortensen S, Greiner M.
\newblock {Relationship of trade patterns of the Danish swine industry animal
  movements network to potential disease spread}.
\newblock Prev Vet Med. 2007 Jul;80(2-3):143--165.
\newblock Available from:
  \url{http://www.sciencedirect.com/science/article/pii/S0167587707000384}.

\bibitem{bigras-poulin_network_2006}
Bigras-Poulin M, Thompson RA, Chriel M, Mortensen S, Greiner M.
\newblock Network analysis of {Danish} cattle industry trade patterns as an
  evaluation of risk potential for disease spread.
\newblock Prev Vet Med. 2006 Sep;76(1--2):11--39.
\newblock Available from:
  \url{http://www.sciencedirect.com/science/article/pii/S0167587706000778}.

\bibitem{Christley:2005}
Christley R, Robinson SE, Lysons R, French NP.
\newblock {Network analysis of cattle movement in Great Britain}.
\newblock In: Proc. Soc. Vet. Epidemiol. Prev. Med.; 2005. p. 234--244.

\bibitem{Dutta:2014hx}
Dutta BL, Ezanno P, Vergu E.
\newblock {Characteristics of the spatio-temporal network of cattle movements
  in France over a 5-year period}.
\newblock Prev Vet Med. 2014 Nov;117(1):79--94.
\newblock Available from:
  \url{http://dx.doi.org/10.1016/j.prevetmed.2014.09.005}.

\bibitem{Rautureau:2012}
Rautureau S, Dufour B, Durand B.
\newblock Structural vulnerability of the French swine industry trade network
  to the spread of infectious diseases.
\newblock Animal. 2012 7;6:1152--1162.
\newblock Available from:
  \url{http://journals.cambridge.org/article_S1751731111002631}.

\bibitem{Webb:2005fs}
Webb CR.
\newblock {Farm animal networks: unraveling the contact structure of the
  British sheep population}.
\newblock Prev Vet Med. 2005 Apr;68(1):3--17.
\newblock Available from:
  \url{http://linkinghub.elsevier.com/retrieve/pii/S0167587705000073}.

\bibitem{Valdano:2015jt}
Valdano E, Poletto C, Giovannini A, Palma D, Savini L, Colizza V.
\newblock {Predicting Epidemic Risk from Past Temporal Contact Data.}
\newblock PLoS Comput Biol. 2015 Mar;11(3):e1004152.
\newblock Available from:
  \url{http://dx.plos.org/10.1371/journal.pcbi.1004152}.

\bibitem{Natale:2009p6279}
Natale F, Giovannini A, Savini L, Palma D, Possenti L, Fiore G, et~al.
\newblock {Network analysis of Italian cattle trade patterns and evaluation of
  risks for potential disease spread}.
\newblock Prev Vet Med. 2009 Jan;92:341--350.

\bibitem{Natale:2011ch}
Natale F, Savini L, Giovannini A, Calistri P, Candeloro L, Fiore G.
\newblock {Evaluation of risk and vulnerability using a Disease Flow Centrality
  measure in dynamic cattle trade networks}.
\newblock Prev Vet Med. 2011 Feb;98(2-3):111--118.
\newblock Available from:
  \url{http://dx.doi.org/10.1016/j.prevetmed.2010.11.013}.

\bibitem{Stark:2006hw}
St{\"a}rk KDC, Regula G, Hernandez J, Knopf L, Fuchs K, Morris RS, et~al.
\newblock {Concepts for risk-based surveillance in the field of veterinary
  medicine and veterinary public health: review of current approaches.}
\newblock BMC Health Serv Res. 2006;6:20.
\newblock Available from:
  \url{http://eutils.ncbi.nlm.nih.gov/entrez/eutils/elink.fcgi?dbfrom=pubmed&i%
d=16507106&retmode=ref&cmd=prlinks}.

\bibitem{Konschake:2013js}
Konschake M, Lentz HHK, Conraths FJ, H{\"o}vel P, Selhorst T.
\newblock {On the Robustness of In- and Out-Components in a Temporal Network}.
\newblock PLOS ONE. 2013 Feb;8(2):e55223.
\newblock Available from:
  \url{http://dx.plos.org/10.1371/journal.pone.0055223.s003}.

\bibitem{HI-Tier}
{Bayerisches Staatsministerium f{\"u}r Ern{\"a}hrung, Landwirtschaft und
  Forsten (StMELF)}.
\newblock {Herkunftssicherungs und Informationssystem f{\"u}r Tiere} [website];
  2016 [cited 2016].
\newblock Available from: \url{www.hi-tier.de}.

\bibitem{RevModPhys.74}
Albert R, Barab{\'a}si AL.
\newblock {Statistical mechanics of complex networks}.
\newblock Rev Mod Phys. 2002 Jan;74:47--97.
\newblock Available from:
  \url{http://link.aps.org/doi/10.1103/RevModPhys.74.47}.

\bibitem{Dorogovtsev:2001jd}
Dorogovtsev S, Mendes J, Samukhin A.
\newblock {Giant strongly connected component of directed networks}.
\newblock Phys Rev E. 2001 Jul;64:025101(R).
\newblock Available from:
  \url{http://link.aps.org/doi/10.1103/PhysRevE.64.025101}.

\bibitem{Newman2003}
Newman MEJ.
\newblock {The Structure and Function of Complex Networks}.
\newblock SIAM Rev. 2003;45(2):167--256.

\bibitem{Lentz:2012pre}
Lentz HHK, Selhorst T, Sokolov IM.
\newblock {Spread of infectious diseases in directed and modular metapopulation
  networks}.
\newblock Phys Rev E. 2012 Jun;85:066111.
\newblock Available from:
  \url{http://link.aps.org/doi/10.1103/PhysRevE.85.066111}.

\bibitem{Newman:2003p4336}
Newman MEJ.
\newblock {Mixing patterns in networks}.
\newblock Phys Rev E. 2003 Jan;67:026126.
\newblock Available from:
  \url{http://link.aps.org/doi/10.1103/PhysRevE.67.026126}.

\bibitem{Lentz:2011}
Lentz HHK, Konschake M, Teske K, Kasper M, Rother B, Carmanns R, et~al.
\newblock {Trade communities and their spatial patterns in the German pork
  production network}.
\newblock Prev Vet Med. 2011;98(2-3):176--181.

\bibitem{holten:2006}
Holten D.
\newblock Hierarchical Edge Bundles: Visualization of Adjacency Relations in
  Hierarchical Data.
\newblock Visualization and Computer Graphics, IEEE Transactions on. 2006
  Sept;12(5):741--748.

\bibitem{Newman:2002p4545}
Newman MEJ.
\newblock {Assortative Mixing in Networks}.
\newblock Phys Rev Lett. 2002 Jan;89(20):208701.
\newblock Available from:
  \url{http://link.aps.org/doi/10.1103/PhysRevLett.89.208701}.

\bibitem{eu_stat_food}
{European Commission}.
\newblock Approved establishments - Lists of approved food establishmentsslide
  [website]; 2016 [cited 2016].
\newblock Available from:
  \url{http://ec.europa.eu/food/food/biosafety/establishments/list_en.htm}.

\bibitem{stat_jahrbuch}
{Bundesministerium f{\"u}r Ern{\"a}hrung und Landwirtschaft (Referat 123)}.
\newblock {Statistisches Jahrbuch {\"u}ber Ern{\"a}hrung, Landwirtschaft und
  Forsten der Bundesrepublik Deutschland}.
\newblock M{\"u}nster-Hiltrup: Landwirtschaftsverl.; 2014.

\bibitem{pastor-sat_2}
Pastor-Satorras R, Vespignani A.
\newblock {Epidemic dynamics in finite size scale-free networks}.
\newblock Phys Rev E. 2002 Mar;65:035108(R).
\newblock Available from:
  \url{http://link.aps.org/doi/10.1103/PhysRevE.65.035108}.

\bibitem{Pastor-Satorras_vespi:2001}
Pastor-Satorras R, Vespignani A.
\newblock {Epidemic dynamics and endemic states in complex networks}.
\newblock Phys Rev E. 2001 May;63:066117.
\newblock Available from:
  \url{http://link.aps.org/doi/10.1103/PhysRevE.63.066117}.

\bibitem{Albert:2000}
Albert R, Jeong H, Barab{\'a}si AL.
\newblock {Error and attack tolerance of complex networks}.
\newblock Nature. 2000;406(6794):378--382.
\newblock Available from: \url{http://dx.doi.org/10.1038/35019019}.

\bibitem{Danon:2011hl}
Danon L, Ford AP, House TA, Jewell CP, Keeling MJ, Roberts GO, et~al.
\newblock {Networks and the epidemiology of infectious disease.}
\newblock Interdiscip Perspect Infect Dis. 2011;2011:284909.
\newblock Available from:
  \url{http://eutils.ncbi.nlm.nih.gov/entrez/eutils/elink.fcgi?dbfrom=pubmed&i%
d=21437001&retmode=ref&cmd=prlinks}.

\bibitem{Lloyd:2001ud}
Lloyd AL, May RM.
\newblock {Epidemiology. How viruses spread among computers and people.}
\newblock Science. 2001 May;292(5520):1316--1317.
\newblock Available from:
  \url{http://eutils.ncbi.nlm.nih.gov/entrez/eutils/elink.fcgi?dbfrom=pubmed&i%
d=11360990&retmode=ref&cmd=prlinks}.

\bibitem{Jones:2003p754}
Jones JH, Handcock M.
\newblock {Social networks: Sexual contacts and epidemic thresholds.}
\newblock Nature. 2003 Jan;423:605.
\newblock Available from: \url{http://www.ncbi.nlm.nih.gov/pubmed/12789329}.

\bibitem{Kao:2006iz}
Kao RR, Danon L, Green DM, Kiss IZ.
\newblock {Demographic structure and pathogen dynamics on the network of
  livestock movements in Great Britain}.
\newblock Proc R Soc B. 2006;273(1597):1999--2007.
\newblock Available from:
  \url{http://eutils.ncbi.nlm.nih.gov/entrez/eutils/elink.fcgi?dbfrom=pubmed&i%
d=16846906&retmode=ref&cmd=prlinks}.

\bibitem{Schwartz:2002dn}
Schwartz N, Cohen R, ben Avraham D, Barab{\'a}si AL, Havlin S.
\newblock {Percolation in directed scale-free networks}.
\newblock Phys Rev E. 2002 Jul;66(1).
\newblock Available from:
  \url{http://link.aps.org/doi/10.1103/PhysRevE.66.015104}.

\bibitem{Meyers:2005p5782}
Meyers L, Pourbohloul B, Newman MEJ.
\newblock {Network theory and SARS: predicting outbreak diversity}.
\newblock J Theor Biol. 2005 Jan;232:71--81.
\newblock Available from:
  \url{http://linkinghub.elsevier.com/retrieve/pii/S0022519304003510}.

\bibitem{Clauset:2009}
Clauset A, Newman MEJ.
\newblock {Power-Law Distributions in Empirical Data}.
\newblock SIAM Rev. 2009;51(4):661.
\newblock Available from:
  \url{http://link.aip.org/link/SIREAD/v51/i4/p661/s1&Agg=doi}.

\bibitem{holme:2002}
Holme P, Kim BJ, Yoon CN, Han SK.
\newblock {Attack vulnerability of complex networks}.
\newblock Phys Rev E. 2002 May;65(5):056109.
\newblock Available from:
  \url{http://link.aps.org/doi/10.1103/PhysRevE.65.056109}.

\bibitem{Pan:2011dga}
Pan RK, Saram{\"a}ki J.
\newblock {Path lengths, correlations, and centrality in temporal networks.}
\newblock Phys Rev E. 2011 Jul;84(1):016105.
\newblock Available from:
  \url{http://link.aps.org/doi/10.1103/PhysRevE.84.016105}.

\bibitem{Holme_review}
Holme P, Saram{\"a}ki J.
\newblock {Temporal networks}.
\newblock Phys Rep. 2012;519(3):97--125.
\newblock Available from:
  \url{http://www.sciencedirect.com/science/article/pii/S0370157312000841}.

\bibitem{Bajardi:2011iv}
Bajardi P, Barrat A, Natale F, Savini L, Colizza V.
\newblock {Dynamical patterns of cattle trade movements.}
\newblock PLOS ONE. 2011;6(5):e19869.
\newblock Available from:
  \url{http://eutils.ncbi.nlm.nih.gov/entrez/eutils/elink.fcgi?dbfrom=pubmed&i%
d=21625633&retmode=ref&cmd=prlinks}.

\bibitem{Casteights_review}
Casteigts A, Flocchini P, Quattrociocchi W, Santoro N.
\newblock {Time-varying graphs and dynamic networks}.
\newblock Int J Parallel Emergent Distrib Syst. 2012;27(5):387--408.
\newblock Available from:
  \url{http://www.tandfonline.com/doi/abs/10.1080/17445760.2012.668546}.

\bibitem{Holme:2015il}
Holme P.
\newblock {Modern temporal network theory: a colloquium}.
\newblock Eur Phys J B. 2015 Sep;88(9):234--30.
\newblock Available from:
  \url{http://link.springer.com/10.1140/epjb/e2015-60657-4}.

\bibitem{lentz:thesis2013}
Lentz HHK.
\newblock Paths for epidemics in static and temporal networks [PhD Thesis].
\newblock Dissertation, Humboldt-University of Berlin; 2013.
\newblock Available from: \url{urn:nbn:de:kobv:11-100213397}.

\bibitem{Rocha_plosbc}
Rocha LEC, Liljeros F, Holme P.
\newblock {Simulated Epidemics in an Empirical Spatiotemporal Network of 50,185
  Sexual Contacts}.
\newblock PLoS Comput Biol. 2011 Mar;7(3):e1001109.
\newblock Available from:
  \url{http://dx.doi.org/10.1371%2Fjournal.pcbi.1001109}.

\bibitem{Valdano:2015il}
Valdano E, Ferreri L, Poletto C, Colizza V.
\newblock {Analytical Computation of the Epidemic Threshold on Temporal
  Networks}.
\newblock Phys Rev X. 2015 Apr;5(2):021005--9.
\newblock Available from:
  \url{http://link.aps.org/doi/10.1103/PhysRevX.5.021005}.

\bibitem{Bajardi:2012}
Bajardi P, Barrat A, Savini L, Colizza V.
\newblock {Optimizing surveillance for livestock disease spreading through
  animal movements}.
\newblock J R Soc Interface. 2012;Available from:
  \url{http://rsif.royalsocietypublishing.org/content/early/2012/06/21/rsif.20%
12.0289.abstract}.

\bibitem{Vernon:2009p5068}
Vernon MC, Keeling MJ.
\newblock {Representing the UK's cattle herd as static and dynamic networks}.
\newblock Proc R Soc B. 2009 Feb;276(1656):469--476.
\newblock Available from:
  \url{http://rspb.royalsocietypublishing.org/cgi/doi/10.1098/rspb.2008.1009}.

\bibitem{Warshall:1962wr}
Warshall S.
\newblock {A theorem on boolean matrices}.
\newblock J ACM. 1962;9(1):11--12.
\newblock Available from:
  \url{http://portal.acm.org/citation.cfm?id=321105.321107}.

\bibitem{Skiena:2008:ADM:1410219}
Skiena SS.
\newblock The Algorithm Design Manual.
\newblock 2nd ed. Springer Publishing Company, Incorporated; 2008.

\bibitem{Grindrod:2011fg}
Grindrod P, Parsons M, Higham D, Estrada E.
\newblock {Communicability across evolving networks}.
\newblock Phys Rev E. 2011 Apr;83(4):046120.
\newblock Available from:
  \url{http://link.aps.org/doi/10.1103/PhysRevE.83.046120}.

\bibitem{Lentz:2013PRL}
Lentz HHK, Selhorst T, Sokolov IM.
\newblock {Unfolding Accessibility Provides a Macroscopic Approach to Temporal
  Networks}.
\newblock Phys Rev Lett. 2013 Mar;110(11):118701.
\newblock Available from:
  \url{http://link.aps.org/doi/10.1103/PhysRevLett.110.118701}.

\bibitem{Nicosia:2012hz}
Nicosia V, Tang J, Musolesi M, Russo G, Mascolo C, Latora V.
\newblock {Components in time-varying graphs.}
\newblock Chaos. 2012 Jun;22(2):023101.
\newblock Available from:
  \url{http://eutils.ncbi.nlm.nih.gov/entrez/eutils/elink.fcgi?dbfrom=pubmed&i%
d=22757508&retmode=ref&cmd=prlinks}.

\bibitem{remark:2014hw}
N{\"o}remark M, Widgren S.
\newblock {EpiContactTrace: an R-package for contact tracing during livestock
  disease outbreaks and for risk-based surveillance.}
\newblock BMC Vet Res. 2014;10:71.
\newblock Available from:
  \url{http://eutils.ncbi.nlm.nih.gov/entrez/eutils/elink.fcgi?dbfrom=pubmed&i%
d=24636731&retmode=ref&cmd=prlinks}.

\bibitem{Euro-Lex2}
{EUR-Lex}.
\newblock {Commission Implementing Decision (EU) 2015/2433 of 18 December 2015
  amending Implementing Decision 2014/709/EU as regards the animal health
  control measures relating to African swine fever in certain Member States}.
\newblock {European Commission}; 2015.
\newblock Available from: \url{http://eur-lex.europa.eu}.

\bibitem{Wang20133869}
Wang Y, Jin Z.
\newblock Global analysis of multiple routes of disease transmission on
  heterogeneous networks.
\newblock Physica A: Statistical Mechanics and its Applications.
  2013;392(18):3869 -- 3880.
\newblock Available from:
  \url{http://www.sciencedirect.com/science/article/pii/S0378437113002756}.

\end{thebibliography}

\begin{thebibliography}{1}
\setcounter{enumiv}{\value{firstbib}}
\bibitem{Grindrod:2011fg_SI}
P.~Grindrod, M.~Parsons, D.~Higham, and E.~Estrada.
\newblock {Communicability across evolving networks}.
\newblock {\em Phys. Rev. E}, 83(4):046120, Apr. 2011.

\bibitem{oie_kitching}
R.~P. Kitching.
\newblock Foot and mouth disease diagnostics: requirements for demonstration of
  freedom from infection.
\newblock In {\em Compendium of technical items presented to the international
  committee or to regional commissions}, pages 189--203, 2001.

\end{thebibliography}

%\begin{thebibliography}{10}
%\include{Lentz_2015.bbl}
%\end{thebibliography}

\clearpage
\appendix
\section*{Supporting Information}

\section*{S1 Assortativity Coefficients}
In Table~\ref{tab:mixing_deg_unweighted} we show the assortativity coefficients for the directed \emph{unweighted} network.
The coefficients were calculated for all combinations between in-degree and out-degree.
For instance in-degree -- out-degree correlation means the tendency that a node of high in-degree has a directed link to a node of high out-degree.
We find that the directed network is still weakly disassortative with respect to the degree.
The relatively high value for out-degree -- in-degree correlations maybe explained by the fact that many small premises trade to nodes of large in-degree, i.e. slaughterhouses or traders.
\begin{table}[htbp]
\caption{Degree assortativity coefficients for the directed unweighted network.}
\begin{center}
\begin{tabular}{lr}
\bf{Correlation} & $r$ \\%& $\sigma$ \\
\hline
in-degree -- in-degree & -0.086  \\
in-degree -- out-degree& -0.063  \\
out-degree -- in-degree & -0.13  \\
out-degree -- out-degree & -0.09  \\
\hline
\end{tabular}
\end{center}
\label{tab:mixing_deg_unweighted}
\end{table}%

\section*{S2 Targeted Vaccination}
In addition to the results in the main text, we compute the size of the GSCC after removing nodes with respect to the following centrality measures:
\begin{description}
%\item[degree centrality.] $C_D$ -- Anzahl der Nachbarn eines Knotens normiert auf die Anzahl der Knoten im Netzwerk.
%Additionally in degree $C_{D^-}$ and out degree $C_{D^+}$.
%\item[betweenness.] $C_B$ -- Häufigkeit, mit der ein Knoten auf kürzesten Verbindungen zwischen anderen Knoten liegt.
%\item[closeness.] $C_C$ -- reziproke mittlere shortest path length eines Knotens zu anderen Knoten.
\item[eigenvector centrality.] $C_E$ -- Correlates with the probability that a node is visited in a random walk on the network.
\item[pagerank.] $C_{P}$ -- similar to $C_E$, but links between arbitrary nodes are possible with a small probability.
\item[Katz centrality.] $C_K$ -- Ability of a node to have shortest path to other nodes, where shorter paths have a stronger weight.
\end{description}

%Table \ref{tab:centr_corr} shows that most correlations are rather small.

%\begin{table*}[htdp]
%\tiny
%\caption{Correlations between centrality measures.}
%\begin{center}
%\begin{tabular}{l|rrrrrrrr}
%\toprule
%{} &  $C_{D^-}$ &  $C_{D+}$ &    $C_D$ &  $C_B$ &  $C_K$ &  $C_C$ &  $C_E$ &  $C_P$ \\
%\midrule
%$C_{D^-}$              &   1.00 &    0.50 &  0.97 &     0.66 &         0.31 &   0.03 &                0.65 &  0.90 \\
%$C_{D^+}$             &    &    1.00 &  0.71 &     0.70 &         0.26 &   0.24 &                0.38 &  0.40 \\
%$C_D$                 &   &     &  1.00 &     0.75 &         0.32 &   0.10 &                0.62 &  0.85 \\
%$C_B$            &    &    &  &     1.00 &         0.25 &   0.06 &                0.50 &  0.59 \\
%$C_K$        &    &     &   &      &         1.00 &   0.16 &                0.22 &  0.32 \\
%$C_C$              &    &     &   &      &          &   1.00 &                0.20 &  0.03 \\
%$C_E$ &    &     &   &      &          &    &                1.00 &  0.62 \\
%$C_P$               &    &     &   &      &          &    &                 &  1.00 \\
%\bottomrule
%\end{tabular}
%\end{center}
%\label{tab:centr_corr}
%\end{table*}%

\begin{figure}[htbp]
   \centering
   \includegraphics[width=\columnwidth]{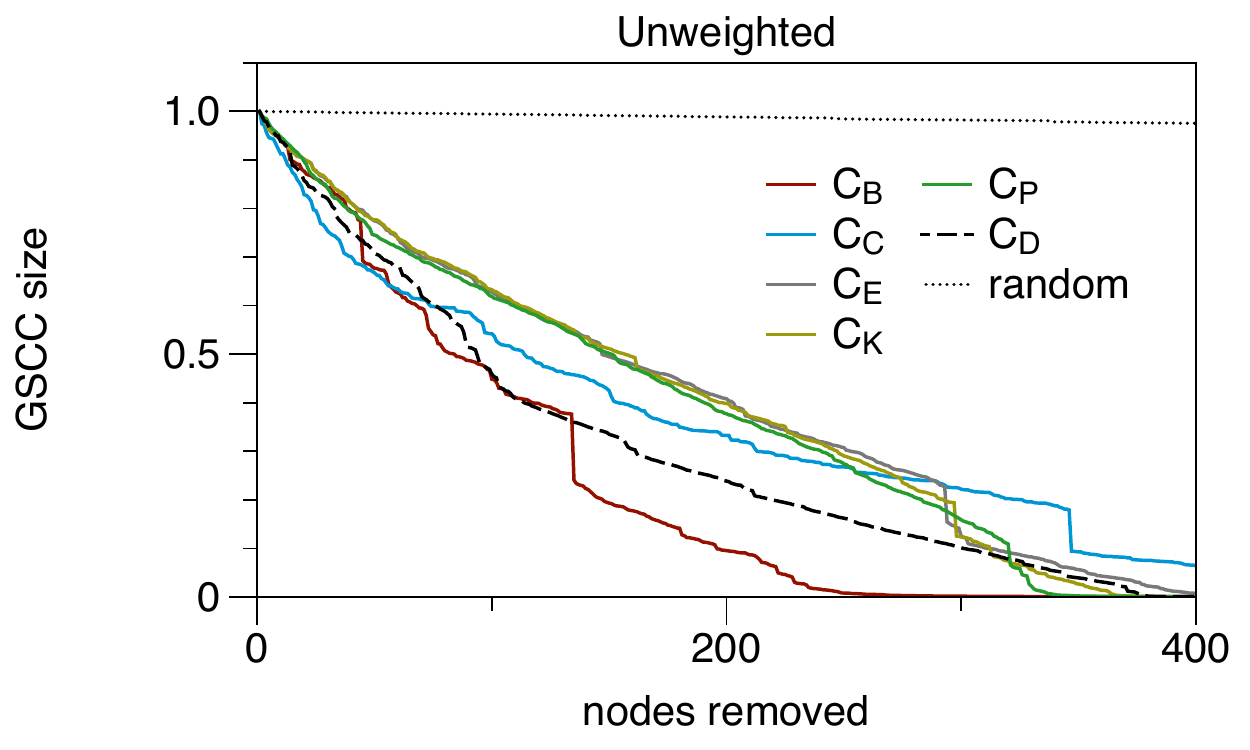} % requires the graphicx package
   \caption{Impact of centrality based node removal, when up to 1 \% of the nodes are removed. $C_B$-Betweenness, $C_C$-Closeness, $C_E$-Eigenvector, $C_K$-Katz-Centrality, $C_P$-Pagerank. Size of giant strongly connected component is normalized to unity.}
   \label{fig:node_removal_SI}
\end{figure}

Figure \ref{fig:node_removal_SI} shows the impact of centrality based node removal for all considered centrality measures.
Although node removal based on eigenvector-centrality, Katz-Centrality or Pagerank still performs better than random node removal, degree betweenness and closeness are more appropriate measures for targeted node removal.

\section*{S3 Centrality in Components}
\begin{figure}[htbp]
   \centering
   \subfloat[Reachability vs. range.]{\includegraphics[width=0.8\columnwidth]{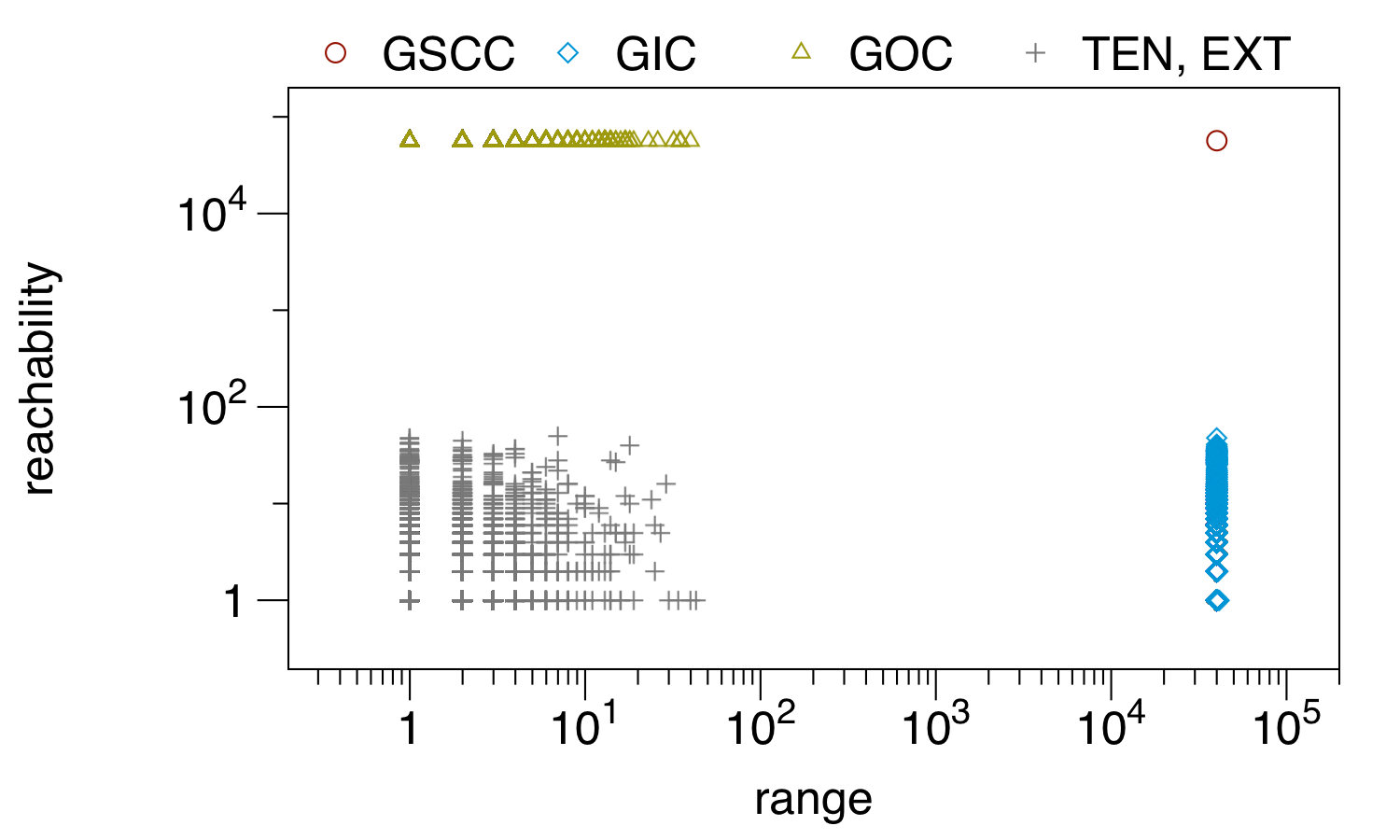}}\\
   \subfloat[Out-degree vs. in-degree.]{\includegraphics[width=0.8\columnwidth]{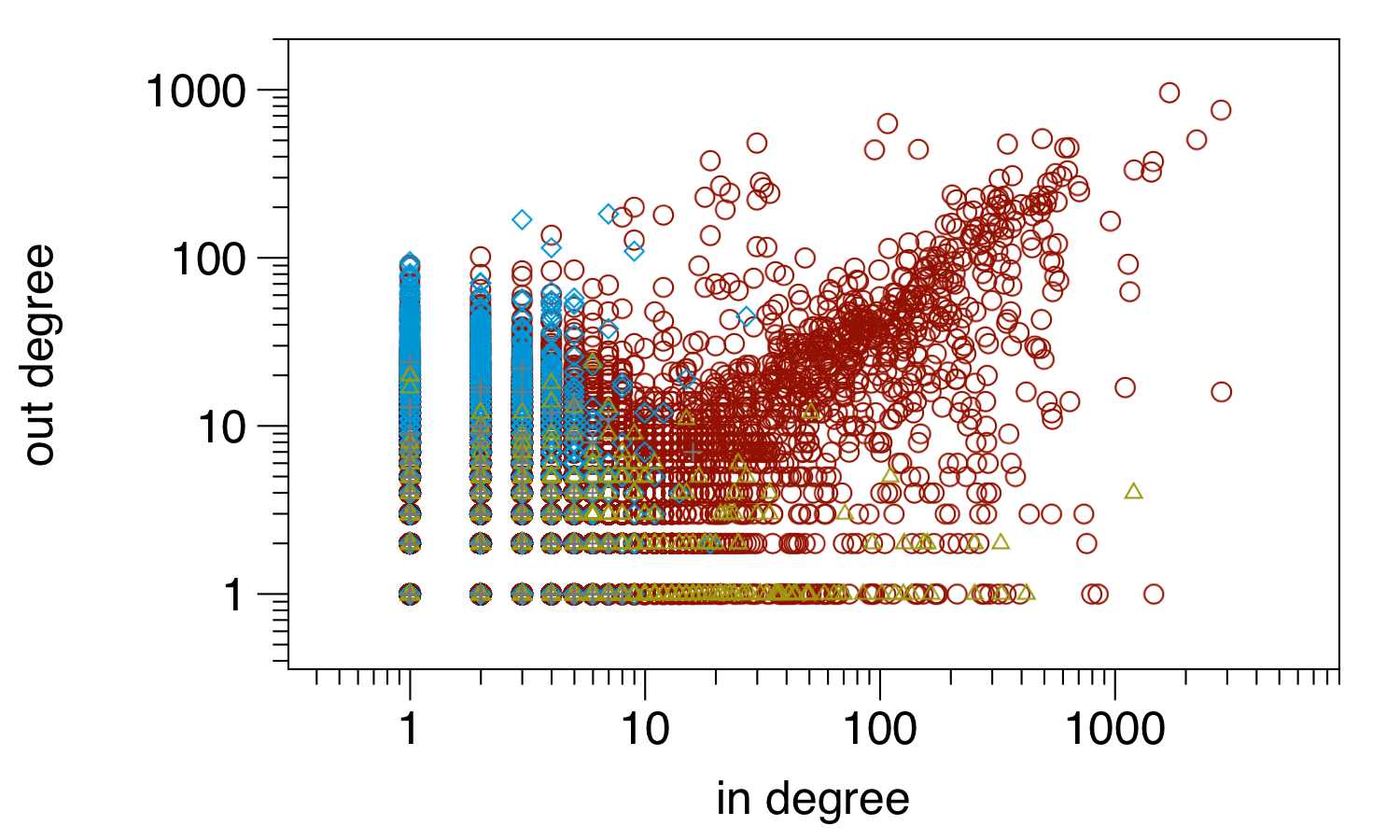}}\\
   \subfloat[Closeness vs. betweenness.]{\includegraphics[width=0.8\columnwidth]{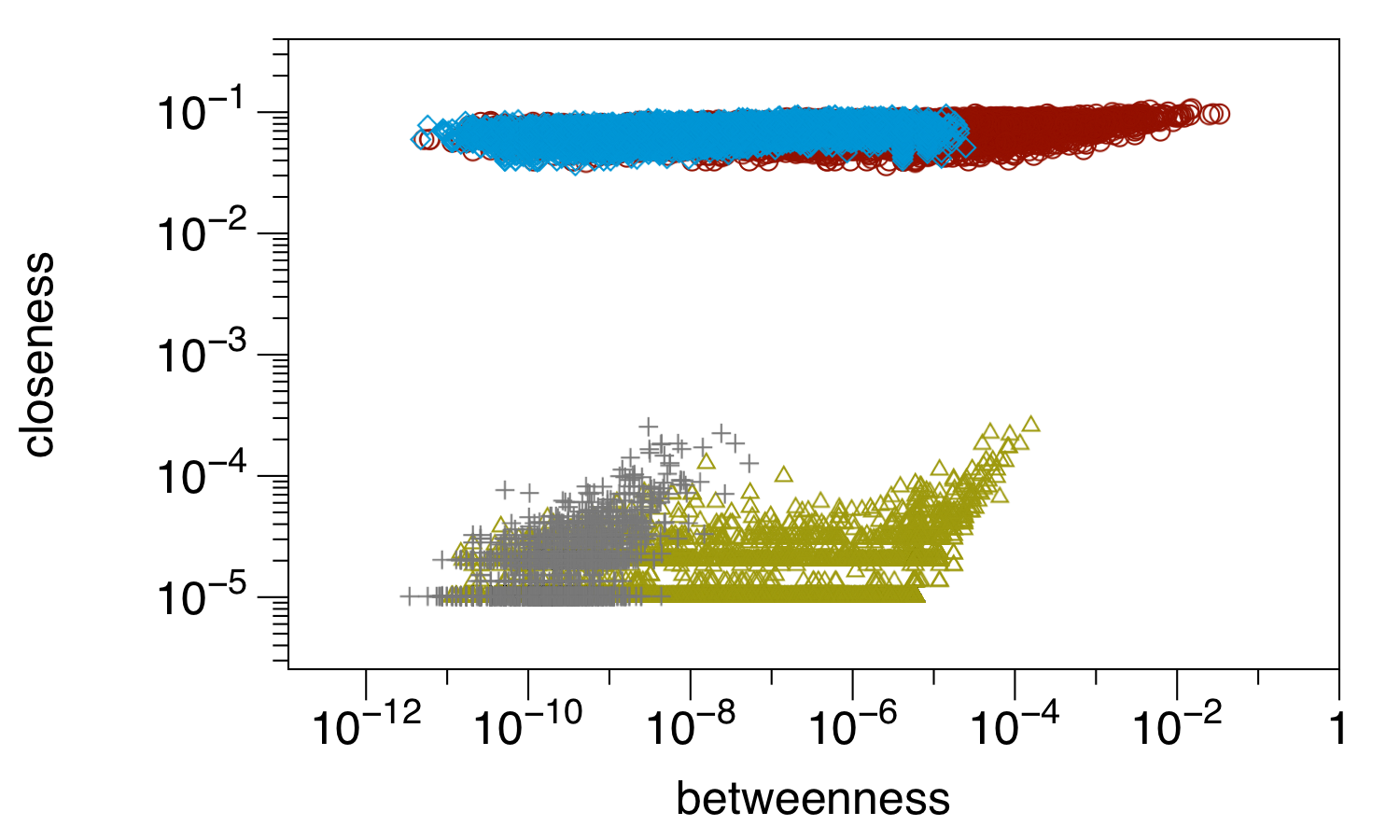}}
   \caption{centrality measures for different giant clusters.}
   \label{fig:centr_clusters}
\end{figure}

In Figure~\ref{fig:centr_clusters} we show the centrality of each node resolved by its giant component membership.
Panel~a) demonstrates that nodes in the GIC and GSCC have a long range.
On the other hand, these nodes have a low reachability.
The reachability of a node is the number of nodes that can reach that node, i.e. its range in the reversed network.
Panel~b) shows that many nodes of high out-degree can be found in the GIC and many high in-degree nodes are located in the GOC.
Furthermore, the correlation between in-degree and out-degree is relatively high in the GSCC.
The GIC and GSCC also contain the nodes of high closeness (Panel~c)).
As expected, nodes with high betweenness are located on the GSCC.

\section*{S4 Weighted Network}
In this section we reproduce the results of the static network in the main text, but take the edge weights into account.
We weight the edges of the network according to the number of traded animals.

First we focus on the large scale structure of the weighted network.
For the component structure, the edge weight does not play any role, since component structure is a purely topological property of the system.
The average shortest path distance is computed using Dijkstra's algorithm in the weighted network.
We find the average shortest path length to be 9.7 and the diameter (longest shortest path length) to be 30.
This implies that weighted shortest paths are on average twice as long as unweighted shortest paths.

%For the weighted link reciprocity \cite{Squartini:2013bk} we find $\varrho = 0.13$.
The assortativity coefficients for the weighted network are shown in Table~\ref{tab:mixing}.
Mixing coefficients for federal state, district and municipality are only marginally influenced by edge weight.
Also the results about the dominant federal states remain similar to the unweighted case:
Inter-state links are mainly formed between North Rhine-Westphalia (NW) and Lower Saxony (NI) as well as Bavaria (BY) and Baden-Wuerttemberg (BW).
These links make up 38 \% of the total trade volume.
The trade between NW and NI alone accounts for 29 \% of all inter-state trade connections.
Concerning the Pareto-principle, 19.3 \% of the weighted edges make up 80.7 \% of all trade volume.

We find that the weighted degree correlations take similar values for all combinations of in-degree and out-degree.
This reflects the fact independent of the combination of degrees, trade is balanced for each pair of premises.
\begin{table}[h]
\caption{Assortativity coefficients between different categories for the weighted network.}
\begin{center}
\begin{tabular}{lr}
\bf{Correlation} & $r$ \\%& $\sigma$ \\
\hline
\hline
Federal state & 0.75 \\%& 0.0007 \\
District & 0.38 \\%& 0.0002\\
Municipality & 0.15 \\%& 0.00006 \\
\hline
degree -- degree  & -0.050  \\
in-degree -- in-degree & -0.041  \\
in-degree -- out-degree & -0.043  \\
out-degree -- in-degree & -0.040  \\
out-degree -- out-degree & -0.042  \\
\hline
\end{tabular}
\end{center}
\label{tab:mixing}
\end{table}%

\begin{figure}[h]
\begin{center}
\includegraphics[width=\columnwidth]{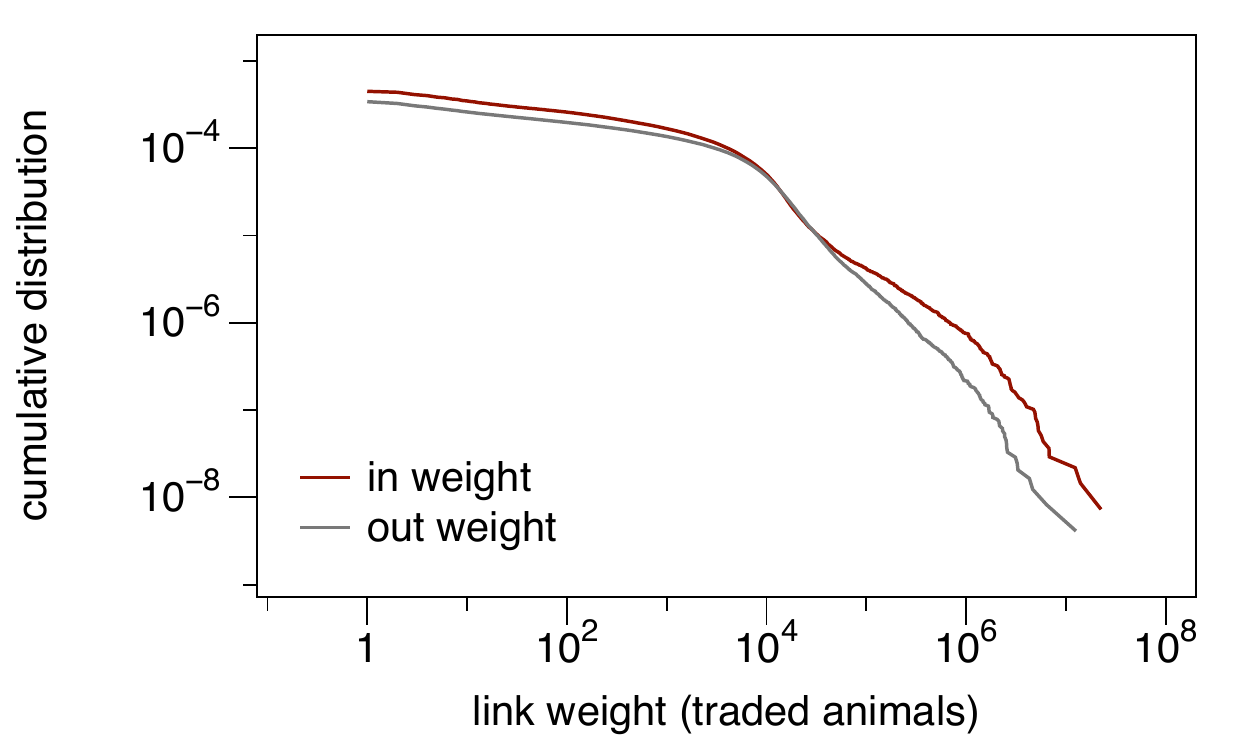}
\caption{Edge weight distribution.}
\label{fig:edge_weight_cdf}
\end{center}
\end{figure}

We now focus on the microscopic structure of the weighted network.
Edge weights can be used in order to compute node centrality more accurately.
The weighted degree distribution shows a similar shape as for the unweighted case.
Figure \ref{fig:edge_weight_cdf} shows the edge weight distribution of the network.
Weight is measured in terms of total number of traded animals during the observation period.

The edge weight plays a significant role for the computation of shortest paths as it is implicitly contained in most centrality measures.
If for instance 1000 animals have been traded from node $i$ to node $j$ and 10 animals have been traded between nodes $i$ and $k$, the weight of the edge $(i,j)$ is significantly higher and this edge would probably be traversed in a shortest path.

\begin{figure}[h]
\begin{center}
\includegraphics[width=\columnwidth]{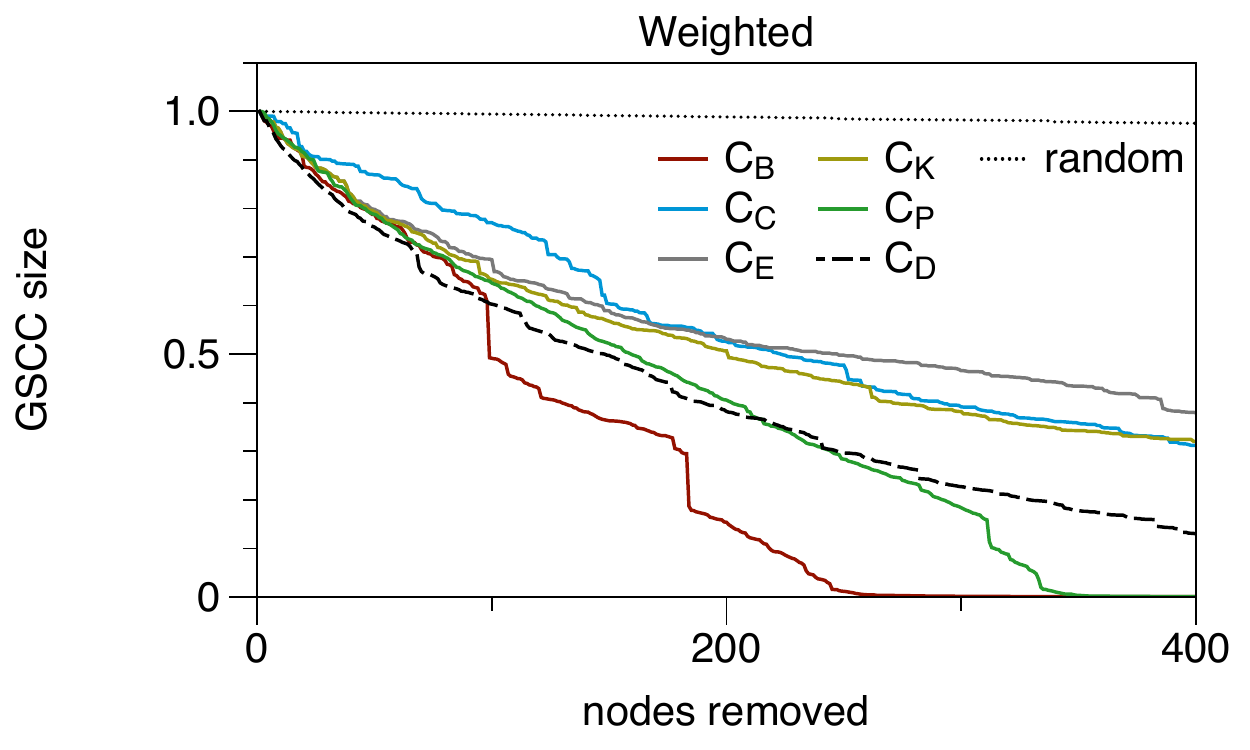}
\caption{Impact of centrality based node removal in the weighted network.}
\label{fig:weighted_attack}
\end{center}
\end{figure}

We compute the centrality measures as in the main text for the weighted network.
Figure \ref{fig:weighted_attack} shows the impact of node removal based on weighted centrality measures.
The results show qualitatively the same behavior as for the unweighted case.

Overall we obtain a similar picture as for the unweighted case:
Nodes of large degree (i.e. sum of trade volume to neighbors) or betweenness perform well for targeted intervention measures.
It is remarkable however that the degree shows a good performance, when only relatively few nodes are removed.
Closeness performs significantly worse than in the unweighted case.

We conclude that:
\begin{enumerate}
\item any centrality based intervention performs significantly better than random intervention!
\item removal of high weight nodes is efficient for removal of up to 100 nodes.
\item removal of high betweenness nodes is efficient for removal of more than 100 nodes.
\item the average edge weight corresponds to a very high infection probability per edge.
\end{enumerate}

\paragraph{Edge Weight vs Infection Probability.}
Finally, we estimate how edge weights can be mapped onto infection probabilities.
Every transport of one or more infectious animals is equally infectious.

First, we compute the probability that \emph{exactly} one animal is infectious $P(X=1)$ for a transport going from premise $i$ to $j$ and $w$ animals are transported.
This probability is given by a binomial distribution $P(X=1) = b_{w,p}(X = 1)$, where $p$ is the probability that an animal is infected in the source node~$i$.
%This probability can be computed using a binomial distribution $B_{w,p}(X\ge 1)$, where $p$ is the probability that an animal is infected in the source node $i$.
This probability is given by the prevalence in node~$i$.

Second, we compute the probability that \emph{at least} one animal is infectious $P(X\ge 1)$ for a transport going from premise $i$ to $j$.
The probability is given by $P(X\ge 1) = B_{w,p}(X = 1)$, where $B_{w,p}(X)$ is the complementary cumulative distribution function (CCDF) of $b_{w,p}(X)$.
The in-farm prevalence for diseases relevant here (classical swine fever, Aujeszky's disease, foot and mouth disease) is typically 30-50~\% at the time of detection \cite{oie_kitching}.

For the data set considered here we observe an average edge weight of $\bar{w} \approx 100$ for every trade transaction (considering the aggregated edge weight would give $\bar{w}_\text{total} \approx 2000$).
Assuming that the prevalence in the source premise is 30~\% as explained above, i.e. $\bar{p} =0.3$, it follows that the probability that an average trade link is infective is given by
\begin{equation}
P(X\ge 1) = B_{\bar{w}, \bar{p}}(X= 1) \approx (100-10^{-14}) ~\% \approx 1,
\end{equation}
where $B_{\bar{w}, \bar{p}}(X)$ is the complementary cumulative distribution function of the binomial distribution.
This means that the expected infection probability per trade transaction is almost 1.
Consequently, the probability of infection if significantly smaller than 1 only for low volume trade transactions.

%\section{Time series}
\section*{S5 Node Activity over Time}
\begin{figure}[htbp]
\begin{center}
\includegraphics[width=\columnwidth]{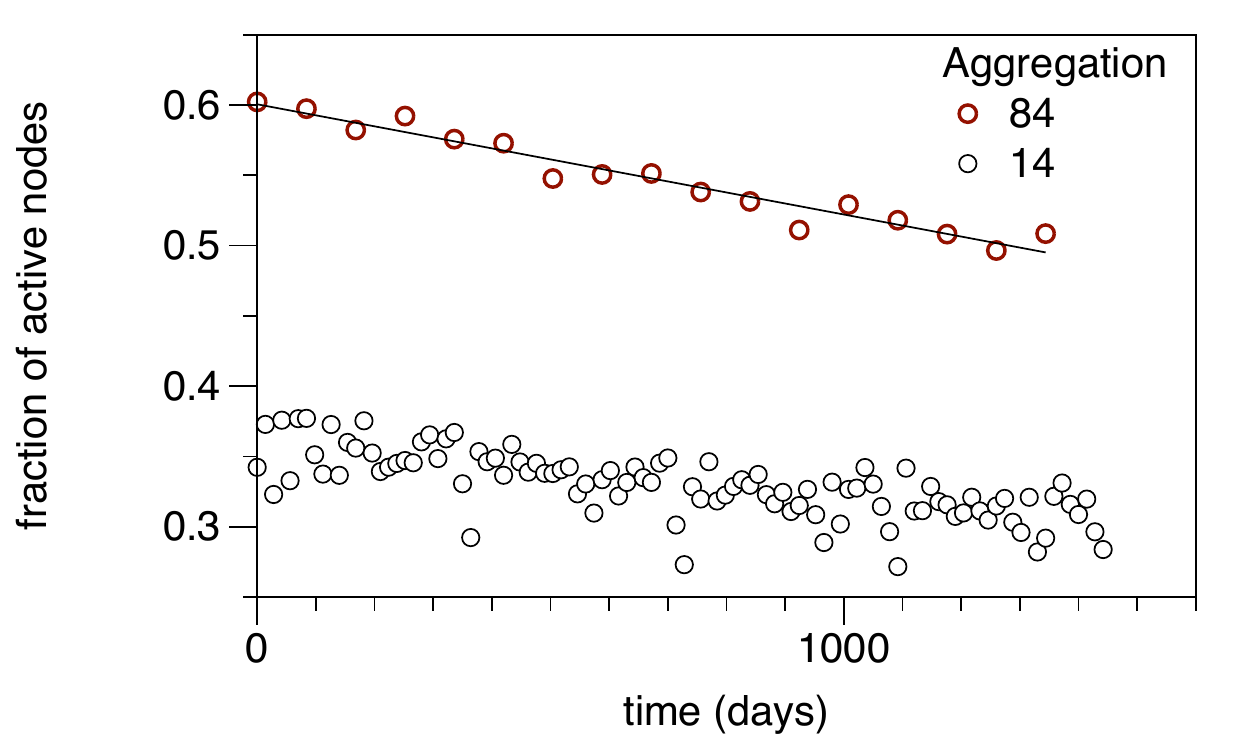}
\caption{Development of the node activity over the observation period.}
\label{fig:node_activity}
\end{center}
\end{figure}

Figure \ref{fig:node_activity} shows the fraction of active nodes over time for two aggregation windows.
For the $84\; d$ aggregation window, the annual loss is approximately 2,800 nodes.
%

%\section{Temporal network}

\section*{S6 Prudent Contact Tracing}
In the tracing procedure mentioned in the main text a pathogen is assumed to only take one step at each snapshot.
There might be necessity for some pathogens that multiple steps are allowed in every snapshot.
In the context of contact tracing, we refer to this circumstance as \emph{prudent} contact tracing.
This corresponds to the situation that a pathogen goes from node $i$ to node $j$ and then from node $j$ to node $k$ and so forth at the same day.
In order to take this into account, we add allow for arbitrary long paths in each snapshot \cite{Grindrod:2011fg_SI}.
Given a temporal network as a sequence of adjacency matrices $\mathcal{A} = \mathbf{A}_1 , \mathbf{A}_2, \dots , \mathbf{A}_T$, we define the long path corrected network as
\begin{equation} \label{eq:prudent_network}
\mathcal{B} = \sum _{i=1} ^{D} \mathbf{A}_1 ^i ,\sum _{i=1} ^{D} \mathbf{A}_2 ^i, \dots , \sum _{i=1} ^{D} \mathbf{A}_T ^i ,
\end{equation}
where $D$ is the diameter of the aggregated network.
The measured value for the diameter is $D=18$ (see main text).
Thus, we allow for a maximum of 18 steps in each snapshot.
This value can be revised downwards depending on assumptions about the disease under consideration or data quality.
Given the path corrected network, prudent contact tracing can be done using the method as described in the main text, but with the temporal network as defined in \eqref{eq:prudent_network}.

\begin{figure}[h]
\begin{center}
\includegraphics[width=0.8\columnwidth]{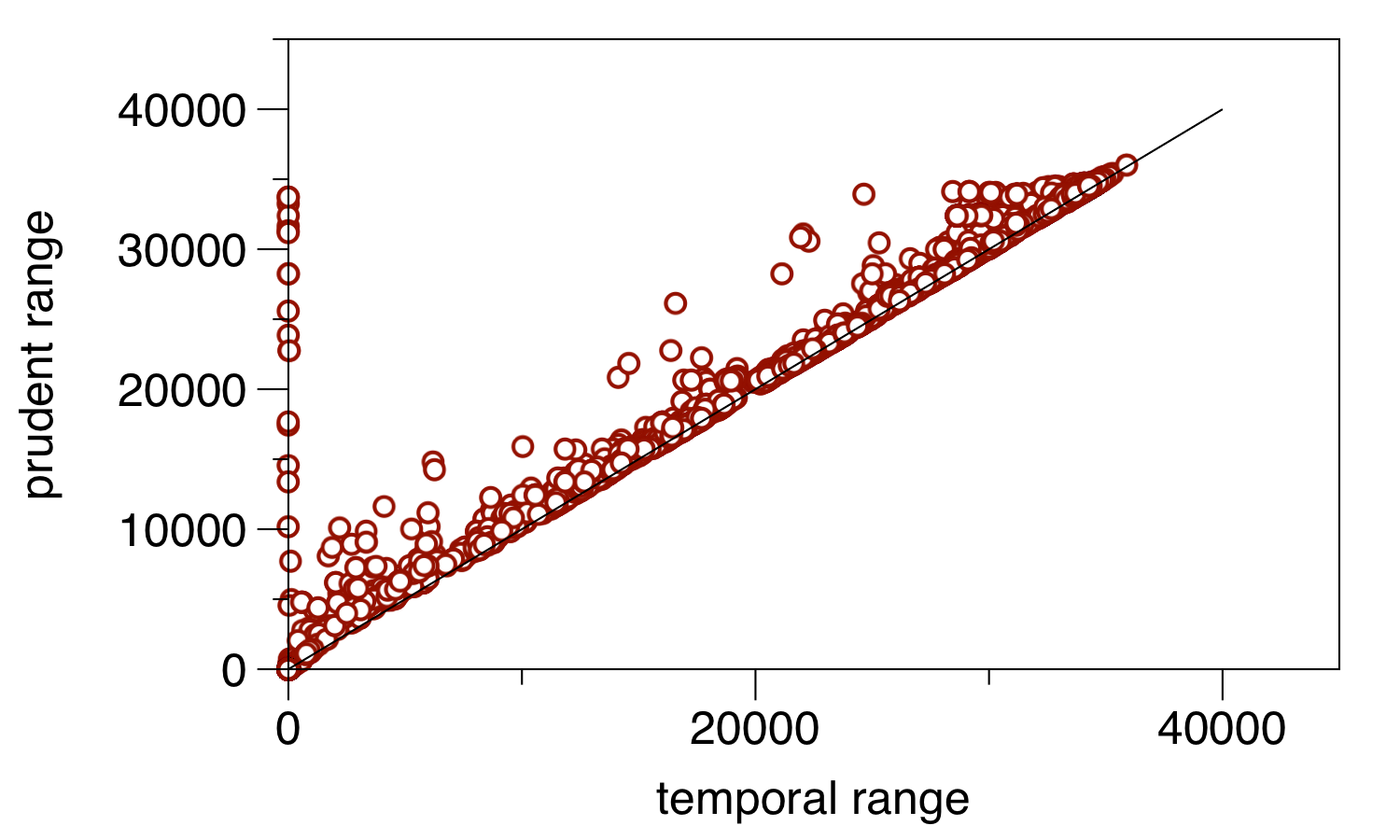}
\caption{Temporal range vs. prudent range for all nodes of the network. Deviation is 148 on average.}
\label{fig:prud_vs_range}
\end{center}
\end{figure}

Figure~\ref{fig:prud_vs_range} shows the range of a node using the standard approach vs. the prudent range computed using \eqref{eq:prudent_network}.
The deviation between the two is 148 on average.

\begin{figure}[h]
\begin{center}
\includegraphics[width=0.8\columnwidth]{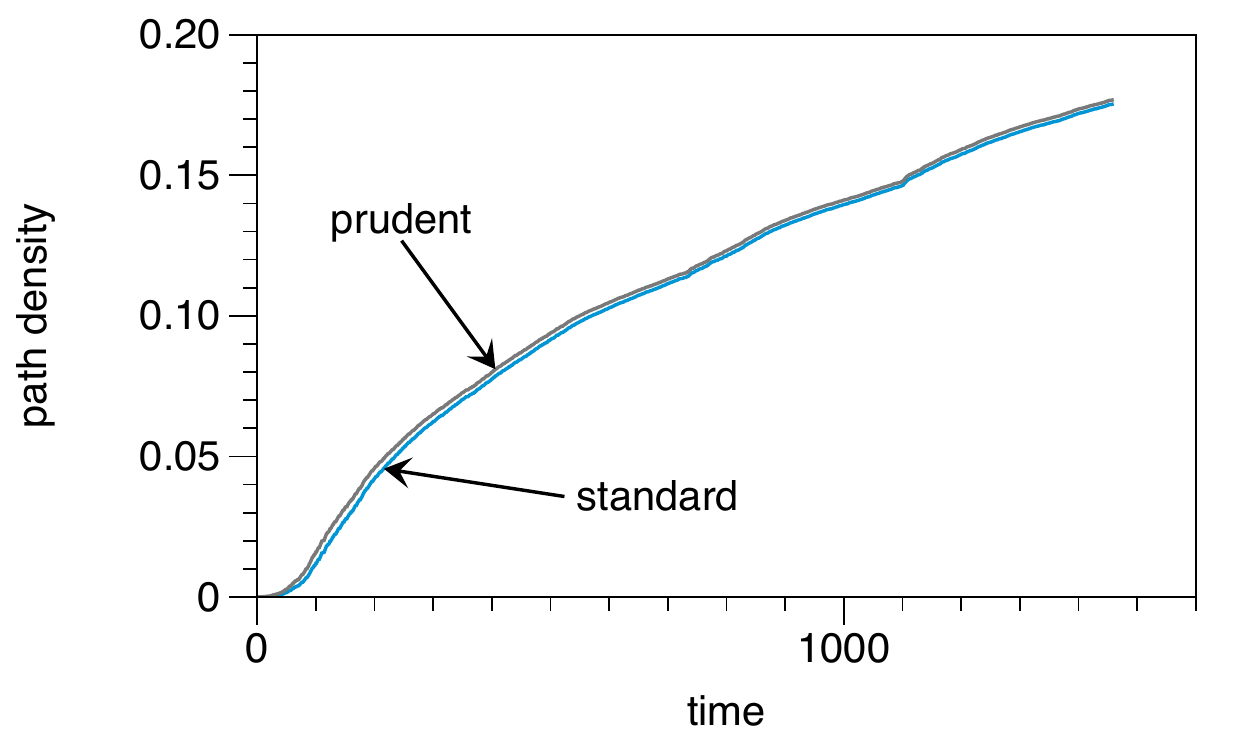}
\caption{Path densities for the network given by $\mathcal{A}$ (standard) and $\mathcal{B}$ (prudent).}
\label{fig:unfolding_prudent}
\end{center}
\end{figure}

Considering the large scale picture of the network, the long path correction does not make a significant difference.
Figure~\ref{fig:unfolding_prudent} shows the path density of the standard approach and the path density computed using  \eqref{eq:prudent_network}.
The curves are almost identical.

\renewcommand{\refname}{References -- Supporting Information}
\newcounter{firstbib}
\setcounter{firstbib}{\value{enumiv}}

%----------------------------------------------------------------------------------------
%\section*{Figure Legends}
%\section*{Tables}

\end{document}